\documentclass[acmsmall,nonacm]{acmart}
\usepackage{algorithm}
\usepackage[noend]{algpseudocode}
\usepackage[probability, asymptotics, primitives, notions]{cryptocode}

\usepackage{hyperref}

\usepackage{float}

\RequirePackage{etex}

\algnewcommand\algorithmicswitch{\textbf{switch}}
\algnewcommand\algorithmiccase{\textbf{case}}
\algnewcommand\algorithmicassert{\texttt{assert}}
\algnewcommand\Assert[1]{\State \algorithmicassert(#1)}%
\newcommand\pks{\mathit{pks}}
\newcommand\pkv{\mathit{pkv}}
\newcommand\skv{\mathit{skv}}
\newcommand\tx{\mathit{tx}}

\newcommand\pkb{\mathit{pkb}}
\newcommand\skb{\mathit{skb}}

\newcommand\owners{\mathit{owners}}
\newcommand\validators{\mathit{validators}}

\newcommand\funds{\mathit{funds}}
\newcommand\settled{\mathit{settled}}

\newcommand\tid{\mathit{tx}}
\newcommand\fid{\mathit{id}}
\newcommand\fbl{\mathit{bl}}
\newcommand\fcert{\mathit{certificate}}

\newcommand\send{\textbf{send }}

\newcommand\bif{\textbf{if }}

\newcommand\upon{\textbf{upon }}

\newcommand\valid{\textbf{valid}}
\newcommand\invalid{\textbf{invalid}}

\newcommand\rcvd{\textbf{received }}
\newcommand\rcpt{\textbf{receipt of }}

\newlength\myindent
\newcommand\whp{{\em w.h.p. }}

\algdef{SE}[SWITCH]{Switch}{EndSwitch}[1]{\algorithmicswitch\ #1\ }{\algorithmicend\ \algorithmicswitch}%
\algdef{SE}[CASE]{Case}{EndCase}[1]{\algorithmiccase\ #1\textbf{:}}{\algorithmicend\ \algorithmiccase}%
\algtext*{EndSwitch}%
\algtext*{EndCase}%
\algnewcommand{\IfThenElse}[3]{
  \State \algorithmicif\ #1\ \algorithmicthen\ #2\ \algorithmicelse\ #3}
\algdef{SE}[WHEN]{When}{EndWhen}%
   [2]{\algorithmicwhen\ \textwhen{#1}\ifthenelse{\equal{#2}{}}{}{(#2)}}%
   {\algorithmicend\ \algorithmicwhen}%
   \algnewcommand\textwhen{\textnormal}
\algnewcommand\algorithmicwhen{\textbf{when}}   
\algtext*{EndWhen}%

\algdef{SE}[EVERY]{Every}{EndEvery}%
   [2]{\algorithmicevery\ \textevery{#1}\ifthenelse{\equal{#2}{}}{}{(#2)}}%
   {\algorithmicend\ \algorithmicevery}%
\algnewcommand\textevery{\textnormal}
\algnewcommand\algorithmicevery{\textbf{every}}   
\algtext*{EndEvery}%

\newcommand{\mc}[1]{}

\newcommand{\remove}[1]{}

\begin{document} 

\title{Breaking the $f+1$ Barrier: Executing Payment Transactions in Parallel with Less than $f+1$ Validations}

\author{Rida Bazzi}
\email{bazzi@asu.edu}
\affiliation{
  \institution{Arizona State University}
  \city{Tempe}
  \state{AZ}
  \country{USA}
}

\author{Sara Tucci-Piergiovanni}
\email{sara.tucci@cea.fr}
\affiliation{
  \institution{Universit\'{e} Paris-Saclay, CEA, List}
  \state{Palaiseau}
  \country{USA}
}

\ccsdesc{Theory of computation~Distributed algorithms}
\ccsdesc{Computer systems organization~Reliability}
\ccsdesc{Security and privacy~Distributed systems security}

\keywords{Distributed Systems, Blockchain, Quorums, Fault Tolerance}

\begin{abstract}
We consider the problem of supporting payment transactions in an asynchronous system in which up to $f$ validators are subject to Byzantine failures under the control of an adaptive adversary. 
It was shown that this problem can be solved without consensus by using byzantine quorum systems (requiring at least $2f+1$ validations per transaction in asynchronous systems). We show that it is possible to validate transactions in parallel with less than $f$ validations per transaction if each transaction spends no more that a small fraction of a balance.
Our solution relies on a novel quorum system that we introduce in this paper and that we call $(k_1,k_2)$-quorum systems. In the presence of a non-adaptive adversary, these systems can be used to allow up to $k_1$ transactions to be validated concurrently and asynchronously but prevent more than $k_2$ transactions from being validated. If the adversary is adaptive, these systems can be used to allow $k_1$ transaction to be validated and prevent more than $k'_2 > k_2$  transactions from being validated, the difference $k'_2-k_2$ being dependent on the quorum system's {\em validation slack}, which we define in this paper. Using $(k_1,k_2)$-quorum systems, a payer can execute multiple partial spending transactions to spend a portion of its initial balance with less than full quorum validation (less than $f$ validations per transaction) then reclaim any remaining funds using one fully validated transaction, which we call a {\em settlement} transaction.\end{abstract}

\maketitle

\section{Introduction}
Existing cryptocurrencies solve the consensus problem to maintain a shared ledger of all transactions, but it was shown that maintaining a shared ledger is not always strictly needed to support exchanging funds~\cite{gupta2016}. In an asynchronous permissioned system in which at most one third of the servers are subject to Byzantine failures, Byzantine quorums can be used to allow different parties to exchange funds through the system -- the asset transfer task. 
The fact that solving consensus is not needed for asset transfer, when the asset has a single owner, was rediscovered by others who formalized the problem, gave it the name {\em asset transfer task} and  generalized it to multi-owner objects~\cite{asset-transfer} as well as generalized the approach to work in a permissionless proof-of-stake system~\cite{abc-chain}. The key insight of~\cite{gupta2016} is that (1) Byzantine quorums can prevent double spending because any two quorums must have at least one correct server in their intersection and (2) ordering unrelated transactions is not needed to exchange funds. Consensus-less solutions are important theoretically, but also in practice for their ability to achieve higher throughput and to reduced transaction latency~\cite{scalable-reliable-broadcast}. 

Despite the increased efficiency of the consensus-less approach, transaction processing is still fundamentally sequential; If one has a balance of ten coins and wants to pay two coins of the ten coins separately to two different recipients, both transactions would be required to be processed by a common correct server to avoid double spending. This, in turn, necessitates every request to be processed by a full quorum so that any two quorums have at least $f+1$ servers in common, where $f$ is an upper-bound on the number of faulty servers. The $f+1$-intersection requirement  ensures {\em safety} so that no two conflicting payments can be simultaneously validated. 

The work on this paper started by considering the possibility of relaxing the intersection requirement so that payments from the same single-owner account could be made 
in parallel while avoiding double spending. 
In the example above, the two single-coin payments are not conflicting, so we would like to allow them to go through without interfering with each other. At the same time, if there were more than ten such single-coin payments, we would like to prevent some of them from going through. These two requirements seem to be conflicting. The question that this paper addresses is the following: 
{\em In an asynchronous system, can we validate non-conflicting asset-transfer transactions in parallel with less than a
full quorum of validators while still being able to prevent double spending? 
}

The surprising answer to this question is in the affirmative, albeit with a some caveats. The main, and provably unavoidable, caveat is that the total balance amount cannot be fully spent with such non-interfering parallel transactions: the owner can make payments in parallel 
up to a given threshold. In order to spend the remainder of the balance, the owner needs to use a larger quorum for validation. In particular, the owner can issue a {\em settlement} transaction to pay the unspent balance to self. The other main caveat is that the solution is probabilistic. Again, this is provably unavoidable. Finally, our current solution requires $n > 8f$ and it is not yet clear how to improve this requirement. An important insight of the solution is captured in the coin example above. If each transaction only spends a fraction of the total balance, then we are able to execute multiple such {\em partial spendings} in parallel, but the total number of such transactions should allow for a buffer between the amount spent and the total balance so that the balance is not exceeded. 
Even more surprising, it turns out that amounts received through partial spending (without full quorum validation) can also be spent partially without full quorum validation if the original payer is correct. If the original payer is not correct, then payees might not be able to spend the amounts received before settlement, but at no time is double spending possible even if all clients are faulty. 

Answering the question in the case of a single owner lead us also to reconsider the {\em proven} need for solving consensus in order to support  transfers from accounts with multiple owners~\cite{asset-transfer}. Intuitively, the need for consensus is to handle double spending from the same account by two different owners. It turns out that partial spending from a multi-owner account is not fundamentally different from the single owner case. Our solution allows multiple owners to spend parts of the balance independently in parallel up to a threshold. The settlement procedure for multi-owner accounts would require solving consensus, which is to be expected given the impossibility result. Our protocols only treat the single owner case.

Our solution is based on a new quorum system that we introduce in this paper and that we call $(k_{1},k_{2})$-quorum system. Unlike traditional quorums in which the main requirement is an intersection requirement, $(k_{1},k_{2})$-quorum systems have both a {\em non-intersection} as well as an intersection requirements. The non-intersection requirement is an upper bound on the size of the overlap between a newly selected quorum and up to $k_1-1$ previously selected quorums. The intersection requirement is a lower bound on the size of overlap between a newly selected quorum and a set of $k_2$ or more previously selected quorums (the actual definition is more subtle and should account for previously corrupted validators).

The solution for executing payment transactions would allow a client using a $(k_{1},k_{2})$-quorum system to execute with high probability $k_1$ partial spending transactions, each of which spends at most $1/k'_{2}$ of the client's balance (for a well specified $k'_2 > k_2$), so the total guaranteed spending is $k_{1}/k'_{2}$ of the clients balance. It is possible that the client can succeed in executing more than $k_{1}$ partial spending transactions but that is not guaranteed. Finally, with high probability the client cannot successfully execute more than $k'_{2}$ partial transactions so the total amount cannot exceed the client's balance. This is achieved even in the presence of corrupt clients that can collude with the servers that validate transactions (the validators).  

The combination of intersection and non-intersection requirements of $(k_{1},k_{2})$-quorum systems is novel and can potentially be applicable to other settings in which we need to allow a limited amount of concurrent activities. In particular, we expect that these systems could be applied to some classes of smart contracts in which available funds significantly exceed actual spending which is done in small increments.
To summarize the main contributions of this paper are the following:

\begin{enumerate}
    \item We introduce $(k_{1},k_{2})$-quorum systems, a new quorum system that has both intersection and non-intersection requirements. The non-intersection requirements allow multiple operations in parallel and the intersection requirements limit the number of operations that can go through without conflict. 
        \item We introduce the {\em partial spending problem} which formalizes the requirements on partial spending transactions and their corresponding settlements.
    \item We present the first protocol that allows multiple payment transactions from the same account to be executed in parallel.
\end{enumerate}

The rest of this paper is organized as follows. Section~\ref{sec:related} discusses related work. Section~\ref{sec:model} presents the system model. Section~\ref{sec:k1k2} introduces $(k_{1},k_{2})$-Byzantine quorum systems. Section~\ref{sec:protocol} presents our solutions. Correctness proofs are presented in Section~\ref{sec:proofs}. Section~\ref{sec:conclude} concludes the paper and discusses how the solution can be simplified for synchronous systems. 

\section{Related Work}\label{sec:related}
 The fundamental bottleneck of  blockchains is the underlying  consensus protocol used to add blocks to the replicated data structure.
 To improve blockchain scalability in the context of cryptocurrencies,  two approaches emerged: 
 asynchronous on-chain solutions that attempt to create  consensus-less blockchain protocols \cite{gupta2016,asset-transfer,abc-chain} and off-chain solutions, as channels and channels factories (e.g. \cite{poon2016, Decker15, Pedrosa19,avarikioti2021b}), that move the transaction load offline while 
resorting to a consensus-based blockchain  only for trust establishment and dispute resolution. 

Asynchronous on-chain solutions rely on the assumption that at most $f$ out of $n > 3f$ servers are Byzantine. In such solutions, payments from the same buyer to different sellers cannot be parallelized and they require at least $f+1$ validations in common (so, at least $2f+1$ validation per transaction). 
As for off-chain solutions, a buyer can set more than one channel to parallelize payments to different sellers, however, all the parties need to agree a priori on the set of participants and  cooperate to operate the channel. Cooperation is needed to open the channel (by locking funds as initial balances), update the state of the channel (by signing transactions that attest of the new allocation of balances) and close the channel by sending the last state update to the blockchain and unlock funds. Frauds are avoided  by constantly monitoring the state of the blockchain, in the case the other party tries to close the channel with a state update different from the last one. Interestingly, to limit cooperation failures and frauds, \cite{avarikioti2021b} uses a set of $n$ processes, called wardens, to proactively validate state updates associated with increasing timestamps agreed by both parties. To get a validation, the buyer needs to contact a full quorum of $t$ of wardens, with $t=2f+1$, under the assumption of at most $f$ out of the $n=3f + 1$ wardens are Byzantine and the non-Byzantine wardens are rational. 
In our setting, we allow a buyer to pay potential sellers in parallel, 
without prior agreement with the sellers and we can support partial spending with validations from less than $f$ validators, all of which can be Byzantine in the case of a corrupt seller.

The closest relative of the $(k_1,k_2)$-quorum systems  that we propose are $k$-quorum systems~\cite{k-quorums,byzantine-k-quorums}. Traditional $k$-quorum systems relax the intersection requirement of quorum systems. A read-quorum is not required to intersect every write-quorum, but they are required to intersect at least one quorum out of any sequence of $k$ successively accessed write-quorums. Successive write-quorums are not required to have a non-empty intersection, but, unlike  $(k_1,k_2)$-quorum systems, there are no requirement that prevents the intersection of two write quorums from being large in size. In systems with Byzantine failures, quorums have more than $f$ elements, and the intersection of a read quorum with the $k$ previous write quorums should be large enough to ensure that some correct servers are in the intersection ~\cite{byzantine-k-quorums}. 
 
Other quorum systems that relate to our proposed quorum system are probabilistic quorum systems~\cite{malkhi1997probabilistic}.

\section{System Model}\label{sec:model}
We consider an asynchronous message passing system of $n$ servers and an unbounded number of clients.  Servers act as {\em validators} for client transactions. Up to $f$ servers and any number of clients can be subject to Byzantine failures under the control of an adaptive adversary. The adversary can corrupt any server up to the limit $f$ and any number of  clients, but the adversary does not have access to the local memory of participants not under its control. If the adversary corrupts a client or server, then the adversary has full control of the corrupted party including the contents of its memory before it is corrupted. We refer to the parties under the adversary's control as corrupt or faulty and to the parties not under the adversary's control as honest or correct. Corrupt parties can deviate arbitrarily from their protocols.
While communication is asynchronous, we make the assumption that messages cannot be selectively delayed to a large enough set of correct servers unknown to the adversary. In other words, if $S$ is a large enough set of correct servers selected by the client's protocol but unknown to the adversary, then any client that sent a request to all servers and is waiting for replies from $n-f$ or $n-2f$ servers, will receive with high probability a reply from an element of $S$. The justification for this is that the adversary can only guess the identities of a few elements of $S$ and selectively delay their communication but cannot do so for all elements of $S$. This is consistent with the goals of our solution whose guarantee should hold with high probability and not in the worst case.

We assume that clients and servers use public-key signature and encryption schemes to sign and encrypt messages and that they are identified by their public keys~\cite{rivest1983method,elgamal1985public}. The public keys of servers are assumed to be known to clients. We assume, but do not explicitly show in the protocols, that messages are signed and signatures are verified. The adversary is computationally bounded and cannot break the encryption or signature schemes. In particular, the adversary cannot read encrypted messages between parties not under its control. Finally,  
parties have access 
to a Hash function in the random oracle model ~\cite{bellare1993random}. 

The adversary can monitor communication between clients and determine which clients are communicating together.
We assume that a client 
can batch messages for multiple transactions together, so that validators for different transactions are contacted together and the adversary cannot tell, just by observing the communication, which of the contacted servers validate which transactions. This point is discussed further when we present the protocols.

\section{The Partial Spending Problem}\label{sec:partial-spending}

\subsection{Payments and Settlements}
Transactions have two clients, the payer and the payee. We refer to them as the {\em buyer} and the {\em seller}, respectively. We use  $\pkb$ to denote a buyer $b$ and $\pks$ to denote a seller $s$.
Spending is done from funds. A fund $F$ is identified by a unique fund identifier $F.\fid$. It has a balance $F.\fbl$, where $\fbl$ is a non-negative amount of money, and one or more owners $F.\owners$ associated with it (in the protocols we only consider single owner funds). Funds are {\em certified} by validators. Each fund has a certificate $F.\fcert$ that consists of a set of validations where each validation has the form $(\langle F \rangle,\sigma,\pkv)$ such that $\sigma = \sign_\skv(\langle F \rangle)$, where $\langle F \rangle$ is an encoding of the fund information (id, balance and owners) and $\pkv$ and $\skv$ are the public and private keys of the validator $v$ that signed the validation. 

Depending on the size of $F.\fcert$, we distinguish between fully certified and partially certified funds. A fund is fully certified if the certificate has validations from at least $f+1$ validators. A fund is partially certified otherwise. Fully certified funds can be initially fully certified, through external means, or are the result of settlement transactions.

To keep the presentation simple, we only consider partial spending transactions from fully certified funds in the formal problem definition and solution and we don't consider partial spending from partially certified funds.
In the appendix, we explain how spending from partially certified funds can be supported.

For fully certified funds, our protocol allows an honest client (using $(k_1,k_2)$-quorums), to execute at least $k_1$ partial spending transactions and at most $k_2$ such transactions, for some global constants $k_1$ and $k_2$ that apply to all partial spending transactions from fully certified funds. In order to avoid double spending by corrupt clients under the control of an adaptive adversary, the amount spent per transaction is $F.\fbl/k'_2$, for some $k'_2 \geq k_2$. The value of $k'_2$ is fully defined as a function of $k_2$, $n$ and $f$.

A {\em partial spending transaction} $\tx:(F,\pkb,\pks, \text{PAY})$ is a transfer of money from fund $F$ with $\pkb \in F.\mathit{owners}$ to a recipient $\pks$. The fund resulting from a transaction $(F,\pkb,\pks, \text{PAY})$ from fully certified fund $F$ is a partially certified fund $F'$ such that $F'.\fid$ is uniquely determined by $F.\fid$, 
$\pks$, $\pkb$ and a payment number $N_s$ which is uniquely generated by the seller for each payment transaction between $\pks$ and $\pkb$ to allow for multiple payments from a buyer to the same seller.
The balance of $F'$ is a fraction of the balance of $F$: $F'.\fbl = F.\fbl/k'_2$, for a global constant $k'_2$, and $F'.\owners = \pks$. The value $1/k'_2$ is the {\em partial spending fraction}.
To keep the presentation simple, we don't support the aggregation of partial spending transactions from separate funds. We note that aggregation can be supported by separate partial spending transactions from separate funds (which can be optimized if the separate funds have the same owners). 
Also, we note here that the model is different from the UTXO model~\cite{nakamoto2008bitcoin} in which no balance remains in the inputs used for payment. In our model,  partial spending transaction leave a balance in the fund from which the payments are made, but the balance is not explicitly maintained. 
In addition to payment transactions, there are {\em settlement transactions}. A settlement transaction $(F,\text{SETTLE})$ specifies a fund $F$ to be settled. The fund being settled can be a fully validated or a partially validated fund, but the fund resulting from the execution of a settlement transaction is fully validated. The identifier of the fund $F'$ resulting from executing  $(F,\text{SETTLE})$ is uniquely determined by the identifier of $F$. 
A validated settlement transaction from fund $F$ results in a fund whose owners are the same as those of $F$ and whose balance is equal to the unspent amount in $F$ and is specified in the safety requirements below. 
\subsection{Problem Definition}\label{sec:problem}
The formal problem definition below only considers requirements for partial spending from initially certified funds (level 1 payments) and settlements associated with those payments. The problem definition does not consider spending from funds resulting from level 1 payments (level 2 payments) or higher level payments that we referred to in the introduction. 
The problem requirements ensure that no double spending is possible. In particular, funds resulting from 
partial spending transactions 
to honest sellers can always be settled 
successfully. The same is not true for funds resulting from partial spending transactions to corrupt sellers. 
In all cases, the total balances of all the funds resulting from settling funds resulting from partial spending transactions from $F$ does not exceed the balance of $F$. 
Finally, we need for honest owners the settlement amount is at least the initial balance minus all payments from the fund (requirement 5); it is not guaranteed to be equal because payment to corrupt sellers could be erased by the adversary. 
These requirements are straightforward to capture.  
Before presenting the requirements, we introduce some notation first. For a given fund $F$, we denote by $\funds_F$ the set of funds resulting from payment transactions of the form $(F,\pkb,\pks, \text{PAY})$ and we denote by $\funds^H_F$  the set of 
such funds for which $\pks$ is honest. 
We denote by $\settled_F$ the set of fully certified funds resulting from transactions of the form 
$(F_{pay},\text{SETTLE})$, where $F_{pay} \in \funds_F$; i.e. the set of settlement transactions of funds resulting from spending money from $F$. 

The following holds with high probability (we use {\em with high probability}, abbreviated \whp, to denote a probability of the form $1-\negl$, where $\negl$ is a negligible function in the relevant parameters):

\begin{enumerate}
    \item (\textbf{progress}) All partially certified funds with honest owners can be settled successfully: Let $F'$ be a partially validated fund resulting from transaction $(F,\pkb,\pks, \mbox{PAY})$ where $F$ is a fully certified fund and $\pks$ is honest. If $\pks$ executes transaction  $(F',\text{SETTLE})$, the transaction will terminate resulting a fully certified fund. 
     \item (\textbf{safety}) If $F''$ is a fully certified fund resulting from executing $(F',\mbox{SETTLE})$ for partially validated fund $F'$, then $F''.\fbl = F'.\fbl$.  

     \item (\textbf{safety}) There can be at most $k'_2$ partial spending transactions from a given fund $F$ for a total spending not exceeding $F.\fbl$.  
    
    \item (\textbf{safety}) Settlement amounts for payments from $F$ are subtracted from settlement for $F$:  If executing transaction $(F,\mbox{SETTLE})$ results in a fully certified fund $F_R$:
   $  F_R.\fbl \leq  F.\fbl - \sum_{F' \in \settled_F} F'.\fbl
   $ 
    
 \item (\textbf{safety}) Payments to honest sellers are subtracted from settlement amount: 
    If executing transaction $(F,\mbox{SETTLE})$ results in a fully certified fund $F_R$: $F_R.\fbl \leq  F.\fbl - \sum_{F' \in \funds^H_F} F'.\fbl$
   
 \item (\textbf{safety}) If the owner of a fund $F$ is honest, settlement amount is no less than the the initial balance of $F$ minus payments made from $F$:  
    If executing transaction $(F,\mbox{SETTLE})$ results in a fully certified fund $F_R$ and $F.\owners$ is honest:
     $  F_R.\fbl \geq F.\fbl - \sum_{F' \in \funds_F} F'.\fbl
    $

    \item (\textbf{progress}) Non-interference: If a total of $k \leq k_1$ payment transactions are initiated by $\pkb \in F.\owners$ from  fully certified fund $F$ and no additional payment or settlement transactions are initiated by $F.\owners$ and $\pkb$ is honest, then,  every one of the $k$ transactions whose seller (payee) is honest will be validated.  
\item (\textbf{progress}) 
Successful settlement for fully certified funds: If the owner of fully certified fund $F$ is honest and executes an $(F,\mbox{SETTLE})$ transaction, the settlement transaction for $F$ will terminate. 
\end{enumerate}

Note that for the non-interference requirement, the guarantee of termination for each transaction holds even if the messages for the remaining transactions are arbitrarily delayed by slow sellers for example. In other words, executing one transaction does not interfere with the completion of other transactions if the total number of transactions does not exceed the threshold $k_1$.

\section{\texorpdfstring{$(k_{1},k_{2})$}--Byzantine quorums}\label{sec:k1k2}
As we explained in the introduction, $(k_{1},k_{2})$-Byzantine quorum systems are what enables our solution.
We first introduce the properties of our new quorum system, then we propose a construction of a particular quorum system that satisfies these properties. In this section, lemmas and theorems are given without proof. Proofs are given in the Appendix. 

\subsection{\texorpdfstring{$(k_1,k_2)$}--quorums: intersection and non-intersection properties}

Our proposed $(k_1,k_2)$-quorum systems have both lower bounds and upper bounds on the sizes and composition of quorum intersections. We discuss their requirements informally before giving the formal definitions. 
In our definitions, we assume, and our protocols ensure, that quorums are selected according according to the uniform access strategy in which all quorum sets are equally likely to be selected. We do not distinguish between read-quorums and write-quorums. 
For the upper and lower bounds on the sizes of intersections, we only consider 
previously corrupted servers, as opposed to servers that are corrupted in response to quorum selection. Previously corrupted servers are chosen randomly according to a distribution that is induced by the probabilistic access strategy. 
Even though we only consider previously corrupted servers in our definition of $(k_1,k_2)$-quorum systems, in our protocols that use these systems, we also consider corruption by an adaptive adversary that might decide on what servers to corrupt based on quorum selection and other information it might have. 
\begin{figure}[t]
  \centering
  \includegraphics[width=2.7in]{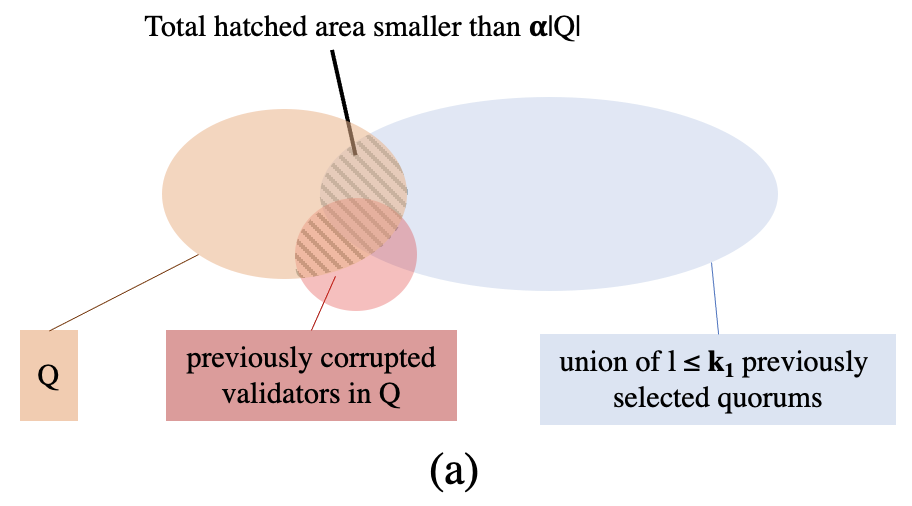} 
    \includegraphics[width=2.7in]{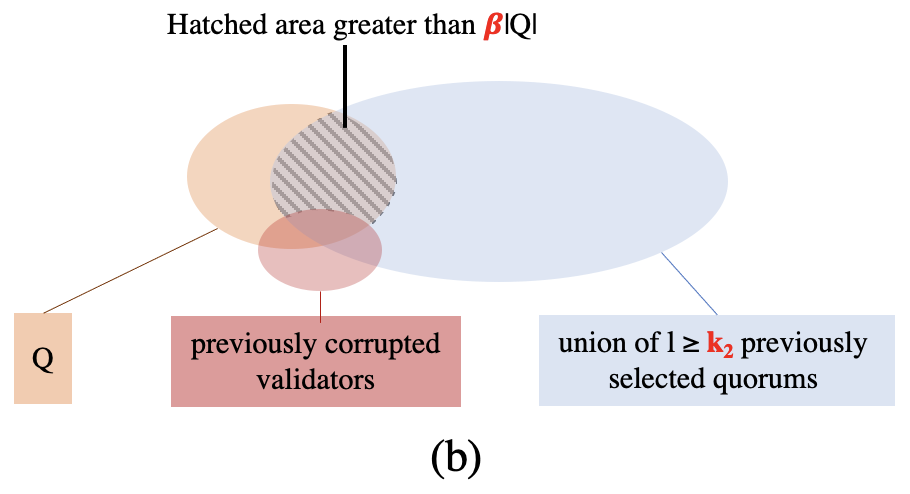}
  \caption{\it Illustration of $(k_1,k2)$-quorums intersection requirements: (a) the intersection of up to $k_1$ previously selected quorums with $Q$ together with previously corrupted servers in $Q$ does not exceed $\alpha |Q|$; (b) the intersection of $k_2$ (or more) previously selected quorums with $Q$ minus any previously corrupted quorums in the intersection is at least $\beta |Q|$;}
  \label{fig:k1k2-illustration}
\end{figure}
 The $k_1, k_2$ parameters are such that the following requirements are satisfied:
\begin{itemize}
    \item (\textit{Non-intersection requirement}): For any newly selected quorum $Q$ and any $\ell \leq k_1$ previously accessed quorums, we require that, with high probability, the size of the union of the previously corrupted servers in $Q$ together with the intersection of $Q$ with the $\ell$ previously selected quorums does not to exceed a fraction $\alpha$ of the size $Q$.     \item (\textit{Intersection requirement}) For any newly selected quorum $Q$ and any $\ell \geq k_2$ previously accessed quorums, we require that, with high probability, the intersection of $Q$ with the union of the $\ell$ previously selected quorums minus the previously corrupted servers in $Q$ has a size that exceeds a fraction $\beta$ of the size of $Q$. In other words, with high probability, the number of correct servers in $Q$ that have been contacted in the previous 
    $\ell$ accesses is more than a fraction $\beta$ of size of $Q$. 
\end{itemize}
The non-intersection and intersection requirements are used to ensure that up to $k_1$ accesses create no conflicts and $k_2$ or more accesses create a conflict. The idea is to allow a transaction to go through if less than $\alpha$ fraction of the quorum have handled potentially conflicting transactions and to deny it if more than $\beta$ fraction of the quorum has handled conflicting transactions. The definition takes into consideration previously corrupted servers who can attempt to coordinate their replies to cause the most damage either to prevent a valid transaction from being validated by claiming to have handled conflicting transactions or to allow an invalid transaction to be validated by validating conflicting transactions. The requirements do not say anything about what holds if $k_1 < \ell < k_2$, which is not an issue for the protocols using these systems. The reason for using two parameters $\alpha$ and $\beta$ instead of only $\alpha$ has to do with the adaptive adversary who can corrupt a randomly chosen quorum after it is chosen. By corrupting selected servers, the adversary can {\em flip} a decision from denying a request to accepting a request. As we will see below, the difference between $\beta$ and $\alpha$ determines how costly it is for the adversary to flip a decision and places an upper bound on the number of such {\em flipped} decisions that the adaptive adversary can achieve.  
In what follows we give the formal definition of $(k_1,k_2)$-quorum systems.

\begin{definition}
A $(k_1,k_2,\alpha, \beta, \epsilon, \delta)$-quorum system is a collection $\mathcal{Q}$ of sets such that:
\begin{enumerate}
    \item Lower bound: $\Pr_{Q , Q_j \leftarrow  \mathcal{Q}\,;\, j \in J_1; \,|J_1| \leq k_1}[|Q \cap ({\mathcal{F}_{pr}}\cup \bigcup_{j \in J_1} Q_j)| > \alpha |Q| ] \geq 1 - \epsilon$
    \item Upper bound: $\Pr_{Q , Q_j \leftarrow  \mathcal{Q}\,;\, j \in J_2; \,|J_2| \geq k_2}[|(Q \cap (\bigcup_{j \in J_2} Q_j))-{\mathcal{F}_{pr}}| \leq \beta |Q| ] \geq 1 - \delta$
\end{enumerate}
where $\mathcal{F}_{pr}$ is the set of faulty servers that existed prior to the random choice of $Q$.
\end{definition}
The lower and upper bound requirements in the definition correspond to the 
non-intersection and intersection requirements stated above and can be best understood by referring to Figure~\ref{fig:k1k2-illustration}.

The definition contains many parameters. The $\epsilon$ and $\delta$ parameters are essentially security parameters that need to be small enough (negligible, in the formal security sense) for the given application. The $\alpha$ and $\beta$ parameters are used for separating the lower and upper bounds. The main parameters from a client point of view are the $k_1$ and $k_2$ parameters, assuming $\epsilon$ and $\delta$ are negligible, hence we refer to these systems as $(k_1,k_2)$-quorum systems.

In $(k_1,k_2)$-quorum systems, individual quorums can be of size less than $f$, which means that even if the randomly chosen quorum does not have any faulty servers, it can be completely corrupted by the adaptive adversary if the adversary knows which servers are in a quorum. In our protocols, this can happen if the seller is corrupt. The definition of the system guarantees that if  no more than $k_1$ accesses are made, less than $\alpha$ fraction of servers in a quorum will deny a request. If more than $k_2$ access are made, then more than $\beta$ fraction of the servers (none of them previously corrupted) will deny the request.  In order for the adaptive adversary to flip the decision, it will need to corrupt $(\beta - \alpha)|Q|$ servers. In fact, by definition, the $\beta|Q|$ servers consisting of previously non-corrupt servers will deny the request with high probability. This number will need to be reduced to no more than $\alpha |Q|$ servers and that would require corrupting $(\beta - \alpha)|Q|$ servers, so that only $\alpha|Q|$ remains. This is the motivation for defining the {\em validation slack}.

\begin{definition}
The {\em validation slack} of a $(k_1,k_2,\alpha, \beta,\epsilon, \delta)$-quorum system  is equal to $(\beta - \alpha)|Q_{min}|$, where $Q_{min}$ is the minimum quorum size
\end{definition}

In total, in a payment protocol using a $(k_1,k_2,\alpha, \beta,\epsilon, \delta)$ quorum system, it is possible for at most $k_2 + f/((\beta - \alpha)|Q_{min}|)$ accesses to be made:  $k_2$ accesses could go through without the intervention of the adversary but the additional $f/((\beta - \alpha)|Q_{min}|)$ accesses can only be achieved by adaptively corrupting validators when clients are corrupt.

\subsection{\texorpdfstring{$(k_1,k_2)$}--quorums Construction}
 In this section, we propose a simple construction of a $(k_1,k_2)$-quorum system to show that they can be constructed. Our construction is not optimal and a more careful choice of the system parameters could yield better performance. Exploring the design space for $(k_1,k_2)$-quorum systems is beyond the scope of this work. 

\begin{definition}
An $(m,n)$ {\em uniform balanced} $(k_1,k_2)$-quorum system is a quorum systems in which each quorum has size $m$, $|U| = n = (k_1+k_2)m$ and quorums are selected uniformly at random.
\end{definition}

\newtheorem*{L1}{Lemma~\ref{lem:uniform-prob}}
\begin{lemma}\label{lem:uniform-prob}
An $(m,n)$ uniform balanced quorum system is a $(k_1,k_2)$-quorum system with $\alpha = 1/3$ and $\beta = 2/3$ has $\epsilon = \delta = \negl[m]$ and {\em validation slack} = $m/3$, if $p_f+\alpha_1 < 1/3$, where $\alpha_1 = k_1m/n$ and $p_f = f/n$.
\end{lemma}

\subsection{\texorpdfstring{Asynchronous $(k_1,k_2)$}--quorum systems}

The calculations and definition of $(k_1,k_2)$-quorum systems implicitly assume that all servers in a quorum reply to a request from a client. 
In order to access a quorum in an asynchronous system, the access should allow for non-responding corrupt servers that cannot be timed out. Assuming that the corrupt servers in a given quorum are randomly chosen,
the expected number of corrupt servers in a quorum of size $m$ is at most $m p_f$, where $p_f = f/n$ is an upper bound on the probability that a randomly selected server has been previously corrupted. With high probability, the number of corrupt servers in a quorum is less than $(1+\mu)m p_f$ for any constant $\mu$, $0 < \mu < 1$, and large enough $m$. In other words, the probability that the number of corrupt servers in a quorum exceeds $(1+\mu) m p_f$ is a negligible function of $m$. It follows that, when contacting a random quorum unknown to the adversary , an honest client can wait for replies from $m(1-(1+\mu) m p_f)$ servers and be guaranteed \whp to receive that many replies. 
Of those replies, $m(1-2(1+\mu) m p_f)$ are guaranteed w.h.p. to be from non-corrupt servers. 

These considerations affect our definition and calculations for $(k_1,k_2)$-quorums in asynchronous systems. The intersection and non-intersection properties should allow for the exclusion of $(1+\mu) m p_f$ servers, which could be correct.  We revise the definition as follows.

\begin{definition}
A $(k_1,k_2,\alpha, \beta, \epsilon, \delta, \mu)$- {\em asynchronous} quorum system is a collection of $\mathcal{Q}$ of sets such that $\forall Q_s \,: |Q_s| \leq (1+\mu) p_f m$, the following holds
\begin{enumerate}
    \item Lower bound:  $ \Pr_{Q , Q_j \leftarrow  \mathcal{Q}\,;\, j \in J_1; \,|J_1| \leq k_1}[|(Q-Q_s) \cap ({\mathcal{F}_{pr}}\cup \bigcup_{j \in J_1} Q_j)| > \alpha |Q| ] \geq 1 - \epsilon$
    \item Upper bound: $
    \Pr_{Q , Q_j \leftarrow  \mathcal{Q}\,;\, j \in J_2; \,|J_2| \geq k_2}[|((Q-Q_s) \cap (\bigcup_{j \in J_2} Q_j))-{\mathcal{F}_{pr}}| \leq \beta |Q| ] \geq 1 - \delta$
\end{enumerate}
where $\mathcal{F}_{pr}$ is the set of faulty servers that existed prior to the random choice of $Q$.
\end{definition}

This definition requires the intersection and non-intersection properties to hold even if we exclude any set $Q_s$ of size up to $(1+\mu) m p_f$ servers from the intersection. Again, the definition has too many parameters. We want $\mu$ and $\alpha$ and $\beta$ to be specific constants for which  $\epsilon$ and $\delta$ are small enough. 
We prove in the appendix the following lemma.

\newtheorem*{L2}{Lemma~\ref{lem:uniform-prob-asynch}}
\begin{lemma}\label{lem:uniform-prob-asynch}
An $(m,n)$ uniform balanced $(k_1,k_2)$-quorum system is a $(k_1,k_2, \epsilon, \delta, \alpha = 1/3, \beta = 2/3, \mu > 0 )$-asynchronous quorum system with $\epsilon = \delta = \negl[m]$ and {\em validation slack} = $m/3$, if $n > 8f$ and $k_1m/n < 1/24$.
\end{lemma}

\section{Solution Overview}\label{sec:overview}
We assume that we have a buyer with a fully certified fund to be spent with partial spending transactions. To understand the difficulties in coming up with a working solution, we start with  a solution that does not work. Then, we successively show how the solution should be modified until we get a working solution. As a first attempt, to make a payment, we can require the buyer to get the transaction validated by a $(k_1,k_2)$-quorum of validators and present the validation to the seller. The idea is that, given the properties of $(k_1,k_2)$-quorum systems, the buyer should not be able to validate more than $k_2$ partial spending transactions and thus double spending is avoided. It turns out that buyer validation is fundamentally flawed due to concerns about settlement and quorum choice. We consider that next. 

{\bf Buyer shouldn't learn the identities of validators:} 
The buyer can always contact the same corrupt quorum for validation which would allow for the validation of more than $k_2$ transactions. To prevent the buyer from choosing the same quorum, we can involve the seller in the choice of the quorum, a choice that should be randomized using the output of the random oracle (hash function) seeded with a combination of 
nonces provided by the seller and the buyer. This would ensure that the choice is random and, if the seller is honest, the buyer has little control over the identities of the validators. Choosing the quorum randomly prevents the buyer from validating more than $k_2$ partial spending transactions, but, 
if the buyer is corrupt, the adaptive adversary would know the identities of the validators and would be able to corrupt them after they are chosen. While the number of transactions that can be validated would still be bounded even if the adversary can corrupt servers in chosen quorums (discussed below), the adversary can create problems when transactions are settled. In fact, consider a fully certified fund $F$ from which a payment is made to an honest seller resulting in partially validated fund $F'$. If $F$ is settled before $F'$ and the validators of $F'$ are corrupted by the adaptive adversary, they can deny knowledge of $F'$ and $F'$ will not be counted against $F$'s balance. This would either result in double spending when $F'$ is settled or in denying the settlement of $F'$ in violation of the problem requirements. This scenario is possible because the buyer knows the identities of the validators and the adversary is adaptive. It follows that the solution should avoid giving the buyer knowledge of the identities of the validators. In our solution, the seller, without involvement from the buyer, chooses the quorum of validators. This works because 
the buyer's and the seller's interests in getting proper validation are at odds. An honest seller has interest in getting as many validators as possible for a given transaction because that would ensure that there is a record of the transaction that they can claim at settlement time. A corrupt buyer has the opposite interest in that it would want to erase any record of a transaction in the hope that the transaction would not count against its balance at settlement time. Of course, the quorum should be randomly chosen and dependent on input from the seller in addition to the transaction identifier. Still this does not solve all the issues. We discuss this next. 

{\bf Protecting the seller during settlement:} 
In the previous section, we discussed how the adversary should not learn the identities of the validators of an honest seller. During settlement, the seller needs to fully validate the settlement transaction and therefore needs to prove to new validators that were not involved in validating the partially validated fund being settled that the fund is properly validated. For that the seller needs to divulge the identities of the validators that validated its partially validated fund. 
The seller cannot send the information to all validators because it is possible for one corrupt validator to receive the information before everyone else. At that point, the adversary learns the identities of the validators and corrupt them to erase their record of the transaction then the buyer can settle its own fund before the seller's settlement request is received by non-corrupt validators, thereby repeating the scenario above.The same scenario can occur during the settlement of a buyer's fund $F$ when validators ncommunicate with each other to determine if there are any partial spending transactions from $F$ that need to be subtracted from the balance owed to the buyer. If the adversary delays the validators' communication except for a pre-chosen corrupt validator, the adversary can learn the identities of all validators that validated partial spendings from $F$ and could then corrupt them to erase the record of some of those partial spendings. 
To prevent this from happening, divulging the identities of validators is done using  secret sharing~\cite{beimel2011secret}. The information 
is divulged in two stages. In a first stage the holder of the information (the {\em dealer}) shares the secret information and in the second stage, the secret information is {\em reconstructed}. 
This way, when the adversary learns the information, it is too late to erase it. 
In our setting, 
we only require that information provided by honest validators can be recovered.

{\bf Buyer should limit number of validators:} 
Since a validator does not validate more than one payment from a given fund, the seller should not get validations from many validators because that can affect the ability of the buyer to get other partial spending transactions validated. The solution to this is to have the buyer approve the validators, but that needs to be done without the buyer learning the identities of the validators. This can be achieved by having the buyer {\em blindly} sign a limited number of {\em validation approvals} such that one approval cannot be used with two different validators. This way, the buyer cannot contact more than a fixed number of validators for a given transaction and the buyer does not learn the identities of the validators.

{\bf The partial spending amount should account for an adaptive adversary and a corrupt seller:}
If the seller is corrupt, then the identities of the validators is known to the adversary, who can then corrupt validators in the selected quorum adaptively. The adversary is limited in the number of validations beyond $k_2$ by the {\em validation slack} which we have discussed earlier. 
We require that if a $(k_1,k_2)$-quorum system with validation slack $s_v$ is used, then the partial spending fraction should not be $1/k_2$ but $1/k'_2$, where $k'_2 = k_2 + f/s_v$.
 
{\bf Corrupt buyer and seller:} If both the buyer and the seller are corrupt, then it is possible to corrupt all the validators in the chosen quorum and erase the record of some payments. This should not create an issue because the amounts that can be erased are only those involving corrupt sellers. Since at settlement time everyone is contacted and a large quorum must validate the settlement,  the buyer and the seller cannot simultaneously receive credit for the erased transaction. 

In what follows, we present the solution and elaborate on each of its components.

\section{Partial Spending and Settlement Protocols}~\label{sec:protocol} 
We start by presenting the protocol for choosing quorums randomly, then we present the partial spending protocol and we finish with the settlement protocol. In what follows we assume that an asynchronous $(k_1,k_2)$-quorum system construction is known to all participants with the parameters $k_1$, $k_2$, $\alpha$, $\beta$, $\mu$ and $m$ available as global constant values.

\subsection{Random Quorum Selection}\label{sec:random}
The $(k_1,k_2)$-quorum systems definition is just a mathematical construction that assumes that a quorum can be chosen randomly, so we need to specify how quorums can be chosen randomly by the seller and how to prevent corrupt sellers from fixing the membership of the chosen quorum.

The protocol (Algorithm~\ref{alg:quorum-select} shown in the Appendix) allows the seller to choose a quorum of $m$ validators and
ensures that if the seller is correct, the quorum of validators is chosen uniformly at random and is not known to the adversary or to the buyer. 
The arguments to the algorithm are a buyer-provided transaction identifier $\tid = \langle F, \pkb, \pks \rangle$ that ties the chosen quorum to the two parties and the specified fund, and a seller-generated random $r$-bit string $N_s$,  where $r$ is a security parameter. The transaction's identifier is concatenated (denoted with $||$) with $N_s$   
to obtain a seller transaction identifier $T_s$, which is 
used as a seed for quorum selection. This seed is 
guaranteed to be unique \whp if the seller is not corrupt. 
The goal is to select $m$ different servers. The seller uses the seed $h$ concatenated with an index $j$ to select the validator $\mathit{Server(}H(h||j))$, where $\mathit{Server}$ is a function that maps the output of $H$ to server identifiers. Since the number of possible validators is $n$, it is possible that the same validator is chosen twice for different values of $j$, so the seller tries successive values of $j$ until $m$ distinct validators are chosen.   
The algorithm returns the identities of the selected validators.  
This information is all that is needed to verify later that a particular quorum of validators was properly chosen according to the protocol. In fact, the set of chosen validators is a deterministic function of the protocol arguments, 
but this requires knowledge of $N_s$ without which an adversary cannot guess the identities of the validators. 

 It is important to point out that the randomness of the quorum is ensured by using $\langle F, \pkb , \pks \rangle ||N_s$ as a seed quorum selection and that a corrupt seller cannot reuse an old seed to double spend because this would result in the same seller transaction identifier.
One final consideration is ensuring random selection for a  corrupt seller that attempts to run the algorithm multiple times in the hope of maximizing the number of previously corrupted validators in the chosen quorum or to present a different $N_s$ at validation time to ensure that the quorum contains a large number of corrupt validators. Since the quorum system is asynchronous, \whp there are no more than $(1+\mu)p_fm$ previously corrupted validators in a randomly chosen quorum. If the seller wants to increase the number to $(1+2\mu)p_fm$, for example, the seller should make an exponential number of attempts (in $\mu p_fm$) to choose a quorum. So, we assume that the quorum system is such that \whp a randomly chosen quorum does not contain more than $(1+\mu/2)p_fm$ previously corrupted validators, and that $(\mu/2)p_fm/2$ is large enough so that \whp the computationally bounded corrupt client cannot chose a quorum with more than $(1+\mu)p_fm$ previously corrupted validators.

The quorum selection protocol satisfies the following properties , which we prove in the appendix.

\newtheorem*{L3}{Lemma~\ref{lem:quorum-random}}
\begin{lemma}\label{lem:quorum-random}
If the seller is correct, the quorum of validators is chosen uniformly at random.
\end{lemma}

\newtheorem*{L4}{Lemma~\ref{lem:unknown-validators}}
\begin{lemma}\label{lem:unknown-validators}
If the seller is correct, the identities of the correct validators in the chosen quorum are not known to the adversary or to the buyer.
\end{lemma}

\newtheorem*{L5}{Lemma~\ref{lem:quorum-corrupt-seller}}
\begin{lemma}\label{lem:quorum-corrupt-seller}
For $0 < \mu < 1$, if a seller choses $K$ quorums, of size $m$ each, at random, then with probability at most $Ke^{-\mu^2 \times p_f m/(2+\mu)}$, every chosen quorums has no more than $(1+\mu)p_fm$ previously corrupt validators.
\end{lemma}

\begin{algorithm}[t]
    \caption{Partial spending: Buyer's Protocol}
    \label{buyer-partial-payment}
    \begin{algorithmic}[1]    
        \Procedure{BuyerPartialSpend}{$\pks$} 
          
            \State $\tid :=  \langle F, \pkb ,\pks \rangle$ \color{blue}\Comment{Transaction Id} \color{black}
            \State \textbf{send} $(\mbox{PAY},\tid)$ \textbf{to} $\pks$ \color{blue}\Comment{send payment message with transaction Id} \color{black}
           
            \State \textbf{wait} \color{blue} \Comment{Wait until commitments}\color{black}
            \State \textbf{until} $(\mbox{QUORUM},\tid,Q_c = [c_i])$ \rcvd \textbf{from} $\pks$ \color{blue}\Comment{ received from seller}\color{black}

                \If{$|Q_c| = m$} \color{blue}\Comment{If quorum has the correct size,}\color{black}

                    \State  \textbf{send} $(\mbox{SIGNED\_QUORUM},\tid,[\sign_\skb(\tid||h_s||c_i)]$) to $\pks$  \color{blue}\Comment{sign and send response}\color{black}

                \Else \color{blue}\Comment{otherwise}\color{black}
                    \State \textbf{return $\bot$} \color{blue}\Comment{abort}\color{black}

                \EndIf
        \EndProcedure
    \end{algorithmic}
\end{algorithm}

\begin{algorithm}[t]
    \caption{Partial spending: Seller's Protocol}
    \label{seller-partial-payment}
    \begin{algorithmic}[1]  
        \Procedure{SellerPartialSpending}{$\tid$} 
            \State $[v_i] = Q_{\tx} \gets \Call{SelectQuorum}{\tid, N_s \leftarrow \{0,1\}^r}$
            \color{blue}\Comment{Seller selects $m$ validators}\color{black}
            \State $[N_i] \leftarrow [(\{0,1\}^r)^m]$ \color{blue}\Comment{$N_i$ is seller's blinding nonce for i'th validator}\color{black}\label{Ns}
            \State $[c_i] \gets \Call{H}{v_i||N_i}$ \color{blue}\Comment{$c_i$ is commitment to i'th validator's identity}\color{black}
            \State $h_s = \Call{H}{N_s}$ \color{blue}\Comment{commitment to $N_s$}\color{black}
            \State $\send (\mbox{QUORUM},\tid,h_s,[c_i]) \textbf{ to } \pkb$ \color{blue}\Comment{send commitments to buyer}\color{black}
        \State \textbf{wait until} $(\mbox{SIGNED\_QUORUM},\tid,[\sigma_i])$ \rcvd \textbf{from} $\pkb$ \color{blue}\Comment{and wait for signatures}\color{black}

        \ForAll{$v \in Q_{tx}$} \color{blue}\Comment{Send signatures with nonces}\color{black}
            \State $\send \langle \tid,h_s,\sigma_i,N_i \rangle \textbf{ to } v$                     \color{blue}\Comment{to validators}\color{black}
        \EndFor
		\State $\mathit{replies} = \mathit{witnesses} = \emptyset$     

        \Repeat 
        \color{blue}\Comment{Wait for replies from }\color{black}
            \If{$\rcvd \mathit{resp} \textbf{ from } v \in Q_{tx} \wedge v \not\in \mathit{replies}$} \color{blue}\Comment{validators while keeping track of}\color{black}
                \State $\mathit{replies} = \mathit{replies} \cup \{\mathit{v}\} $ 
                        \color{blue}\Comment{ which validators replied and}\color{black}

            \If{$\mathit{resp}  = \langle \valid, \tid,h_s, \sigma = \sign_\skv(\tid||h_s)\rangle$} \color{blue}\Comment{which validators}\color{black}
                \State $\mathit{witnesses} = \mathit{witnesses} \cup (\mathit{resp},v) $ \color{blue}\Comment{are witnesses}\color{black}
            \EndIf
            \color{black}
            \EndIf

        \Until{$|\mathit{replies}| \geq  m - (1+\mu) p_f m$} 

        \If{$|\mathit{witnesses}| \geq (1 - \alpha) \times m$}         \color{blue}\Comment{If enough validators validated,}\color{black}
            \State \textbf{return } $\langle \tid, N_s,\mathit{witnesses}\rangle $
        \color{blue}\Comment{return certificate for partial spending transaction}\color{black}
        \Else
            \State \textbf{return $\bot$}
        \EndIf
        \EndProcedure
    \end{algorithmic}
\end{algorithm}

\subsection{Partial Spending Protocols}

The partial spending protocols for the buyer and seller are shown in Algorithms~\ref{buyer-partial-payment} and~\ref{seller-partial-payment} respectively. The validator code relating to partial spending is shown in Algorithm~\ref{validator-payment}. 
The code closely follows the solution overview above. 
In the code we assume that the quorum parameters including the validation slack $v_s$ are fixed and known by all parties and therefore the partial amount to be paid from fund $F$, which is equal to  $F.\fbl/(k_2+f/v_s)$, needs not be explicitly shown in the code.

\subsubsection{Buyer's Code} 
The buyer sends a PAY message specifying the transaction identifier (recall that we only consider partial spending transactions in the protocols). 
The buyer then waits for a response from the seller
which will be a vector of commitments $[c_i]$ to the identities of a quorum of validators (explained below in the overview of the seller's code). 
The buyer checks that the quorum size is $m$ and sends a reply that includes for each validator  a signature of $\tid||c_i$ to link the commitment to the transaction.
The buyer does not need to check that the chosen quorum is valid. That is done when the seller settles the payment. 
\begin{algorithm}[t]
    \caption{Validator $v$ Payment Validation}
    \label{validator-payment}
    \begin{algorithmic}[1] 
     \If{ \textbf{received} $\langle \tid,h_s,\sigma,N \rangle$ \textbf{from} $\pks$ $\wedge \,$  
     $\tid.F \not\in \mathit{settle}$ $\wedge$
     
$(\tid.\pkb \in \tid.F.\owners) \wedge$
$ (\tid.\pks = \pks) \,\wedge$
$(\tid.F \not\in \mathit{validated\_fund})$ $\wedge$ 
     
     \Call{VerifySig}{$\tid.\pkb,\tid||h_s||\textproc{H}(v||N)),\sigma$} }
     \State $\mathit{validated\_fund} = \mathit{validated\_fund} \cup F$
     \State $\mathit{validated\_transactions} = \mathit{validated\_transactions} \cup \langle \tid, h_s,\sigma, N \rangle$
     \State \textbf{send} $\langle \valid, \tid ,h_s, \sign_\skv(\tid||h_s) \rangle$ \textbf{to} $\pks$  
     \Else
     \State \textbf{send} $\langle \invalid, \tid \rangle$
        \textbf{to} $\pks$      
    \EndIf
    \end{algorithmic}
\end{algorithm}

\subsubsection{Seller's Code}
The seller's \textproc{SellerPartialSpending()} function is executed in response to a PAY message from the buyer and 
takes the $\tid$ of the PAY message as argument. The seller starts by calling \textproc{SelectQuorum()} using the buyer's transaction identifier and a randomly generated nonce $N_s$. The call  returns a quorum of validators $Q_\tid$ represented as a vector of validators $[v_i]$.
The seller generates $m$ $r$-bit nonces that it uses in generating a vector of commitments  
$[c_i] = [H(v_i||N_i)]$ to the validators identities, which it sends to the buyer, and waits to receive from the buyer the signed commitments. Then the seller sends to each validator a validation request that includes the transaction identifier $\tid$, the signature $\sigma_i$ (which should be equal to 
$\sign_\skb(\tid||c_i)$), and the nonce $N_i$ used in generating the commitment. It then waits to receives replies from  $m-(1+\mu)p_fm$ validators 
and checks if $(1-\alpha) m$ validators validated the transaction. If so, the transaction is validated and the replies from validators that validated the transaction constitute a {\em certificate} of validation for the transaction. In the code, we do not explicitly represent the resulting fund, but we note  that the transaction, and therefore the fund that it creates, is specified by
the transaction identifier $\tid = \langle F, \pkb, \pks \rangle$ and $hs$, the seller's commitment to the nonce $N_s$.

\subsubsection{Validator's Code}
The validator checks for the validity of the request and sends the result of the validation to the seller. A request is valid if: the fund from which the payment is made has not been settled; the seller specified by the transaction is the one making the request; the buyer specified in the transaction is the buyer specified in the fund of the transaction; the fund specified by the transaction has not been previously validated by the validator; the signature 
$\sigma$ in the request is a valid signature for $\tid||c$ by the buyer $\tid.\pkb$ specified in the request; and, $c$ is a valid commitment to the validator's identity.
If the request is valid, the validator adds the fund to the set of validated funds and the transaction to the list of validated transactions. The list of validated funds is needed to ensure that a validator validates only one partial payment from a given fund. It is used during settlement of buyer's fund $F$ to ensure that all partial spending from $F$ will be accounted for. That is why $\sigma$, $c$ and $N$ which are needed to validate a transaction are stored along with $\tid$.

\subsection{Settlement Protocols}\label{sec:settle}
As we explained in the solution overview, during settlement there is a need to prevent the adversary from learning the identities of the validators prematurely and corrupting them. 
The settlement algorithms use a \textproc{Propagate()} protocol that allows a party to propagate information to all but $2f$ correct servers without the adversary having the ability to suppress the information being propagated. We outline the \textproc{Propagate()} protocol first (the details and code are in the appendix), then we present the settlement protocols for the seller, buyer and validators.

\subsubsection{Propagating Information}
As we discussed in the protocol overview, the adversary can 
corrupt a validator if it learns that the validator validated a 
particular transaction and the adversary wants to suppress that information. The identities of validators that are involved 
in validating a particular transaction are kept secret by the validators themselves (if honest) or the seller until settlement time.  The \textproc{Propagate()} protocol allows a  party (seller or validator) to send a message to all validators so that the adversary would either
have to corrupt the party to learn the message or, if the adversary learns the message without corrupting the party, then all but $2f$ honest are also guaranteed to learn the message. The \textproc{Propagate()} protocol is implemented using secret sharing and reconstruction.  
To distinguish between messages propagated by different calls to \textproc{Propagate()} by clients, a unique $N_{prop}$ nonce is generated for the call and provided as a second argument to \textproc{Propagate()}. When a message from client $c$ is finally propagated to a server, the server will have $\mathit{message}[c,N_{prop}]$ equal to the propagated message. The details are given in the appendix.

\subsubsection{Seller's Settlement and Corresponding Validation} The settlement algorithm for the seller is straightforward. The seller  propagates a settlement request to validators. The fund is fully specified by $\tid = \langle F, \pks,\pkb \rangle$ and $N_s$ of the partial spending transactions that created the fund.
The certificate of the fund consists of the set of $\mathit{payment\_witnesses}$ that validated the partial spending transaction. The information $\tid, N_s, \mathit{payment\_witnesses}$ enables a validator of the settlement transaction to recalculate the quorum used in validating the transaction and to verify that the witnesses
are provided by members of that quorum. The validator checks that the quorum and validations provided by the seller are correct and that the validation for the transaction are received from a large enough subset of validators. In addition, the validator checks that either it has no record of SETTLE transaction for $F$ ($F \not\in \mathit{settle}$) or the transaction being settled, represented as $(\tid, h_s = \Call{H}{N_s})$ was added to $\mathit{transactions}[F]$ which is the set of transactions calculated when the buyer settles $F$. This check is needed to avoid double spending. As shown in the proofs, if the buyer settles $F$, every payment to an honest seller will be added to  
$\mathit{transactions}[F]$ which ensures that this condition will be satisfied for honest sellers.
If all the information checks out, the validator sends a validation for the settlement. The seller waits until it receives $n-f$ validations, which is guaranteed to happen if the seller is correct. These $n-f$ validations guarantee that only one fund can result from settling a partially certified fund. Finally, note that the settlement of the seller's partially certified fund  can go through even the validators that validated the original spending transaction are corrupted.

\begin{algorithm}[t]
    \caption{Seller Fund Settlement}
    \label{seller-settlement}
    \begin{algorithmic}[1] 
        \Procedure{SellerSettle}{$\tid = \langle F, \pkb,\pks \rangle, N_s,\mathit{payment\_witnesses}$}
       \State $N_{settle} \leftarrow \{0,1\}^r$
     \State \Call{Propagate}{$\langle \tid, N_s, \mathit{payment\_witnesses},\text{SETTLE}\rangle$, $N_{settle}$} 
        \Repeat 
            \If{$\rcvd (\textbf{valid},\sigma, N_{settle} )$ \textbf{ from } $v$}
            \State $\mathit{Id}_1 = \Call{H}{\tx||N_s||\mbox{PAY}}$;
            \hspace{3ex} $\mathit{Id}_2 = \Call{H}{\mathit{Id}_1||\mbox{SETTLE}}$
            \State $F'.\fid = Id_2 \,;\,  F'.\fbl =  F.\fbl/k'_2 \,;\, F'.\owners = \{\pks\}$
            \If{$\sigma=\sign_v(\langle F'.\fid, F'.\fbl, F'.\owners\rangle$}
                \State $\mathit{settle\_witnesses} = \mathit{settle\_witnesses} \cup  \{(v,\sigma)\}$
            \EndIf
            \EndIf
        \Until{$|\mathit{settle\_witnesses}| \geq n-f$}
        \State $F'.\fcert = \mathit{settle\_witnesses}$
        \State $\textbf{return } (\langle F'\rangle)$
       \EndProcedure
    \end{algorithmic}
\end{algorithm}

\begin{algorithm}[t]
    \caption{Code of Validator $v$ for Seller Fund Settlement}
    \label{validator-seller-settlement}
    \begin{algorithmic}[1] 
           \If{$\mathit{message}[\pks,N_{settle}] = \langle \tid, N_s, \mathit{payment\_witnesses},\text{SETTLE}\rangle$  $\wedge$ 
       
       $Q = \Call{SelectQuorum}{\tid, N_s}$ $\wedge$  
		$((\tid, \Call{H}{N_s}) \in \mathit{transactions}[F] \vee F \not\in \mathit{settle})$ $\wedge$
       
       $\Call{ValidatedTransaction}{\tid,\mathit{payment\_witnesses}}$ $\wedge$  
       
       $Q' = \{ q\,:\, (\mathit{resp},q) \in \mathit{payment\_witnesses}\} \subseteq Q$ $\wedge$ 
        $|Q'| \geq m-m\mu p_f$ }
            \State $\mathit{transactions}[F] = \mathit{transactions}[F] \cup \{((\tid, \Call{H}{N_s})\}$
       		\State $\mathit{Id}_1 = \Call{H}{\tx||N_s||\mbox{PAY}}$
            \State $\mathit{Id}_2 = \Call{H}{\mathit{Id}_1||\mbox{SETTLE}}$
            \State $F'.\fid = Id_2 \,;\,  F'.\fbl =  F.\fbl/k'_2 \,;\, F'.\owners = \{\pks\}$
            \State \send $(\textbf{valid}, \sign_v(\langle F'.\fid, F'.\fbl, F'.\owners\rangle, N_{settle})$ to $\pks$
            \EndIf
           
    \end{algorithmic}
\end{algorithm}
\subsubsection{Buyer's Settlement and Corresponding Validation} To settle the buyer's fund $F$, we need to make sure that all partial spending from the buyer's fund are deducted from the settled balance. The buyer's starts by sending a settlement request to all validators. Every validator that receives the buyer's request {\em propagates} information about any payments from $F$ to sellers that it is aware of (there can only be at most one such payment per validator). A validator will either provides proof that it witnessed a payment or states that it did not witness any payment. The proof of a witnessed payment $(\tid, \sigma, N_h)$ where $\sigma = \sign_{\pkb}(H(\tid||N_h))$ is the blind signature and $N_h$ is the blinding factor. Validators wait until $n-f$ different messages about payments from $F$ are propagated. Every validator then counts the number of different payment transactions from $F$ that it heard about, calculates the resulting balance after subtracting the amounts for those payments and send the seller a signed balance for the settlement fund (see problem definition). The buyer waits until it receives $n-2f$ signatures for the same balance which will form the set of witnesses for the settlement fund resulting from $F$. The code is given in the appendix.

\section{Conclusion}\label{sec:conclude}
We introduced $(k_1,k_2)$--quorum systems and shown how to use them to execute payment transactions with less than $f$ validations per transaction. By carefully considering intersection and non-intersection properties, we are able to allow up to $k_1$ non-interfering payments in parallel and prevent double spending by limiting the number of such concurrent transactions. It might seem that the construction succeeds because when the seller is honest, the quorum is randomly selected and has, with high probability, a small fraction of corrupt validators. We observe that in the case of corrupt sellers, the whole quorum could be corrupted, and our definition of  $k'_2$ was done explicitly to handle that possibility. 
The constructions we proposed work for large values of $n$. It is not clear how to modify the constructions to work for smaller values of $n$ and larger values of $f$ relative to $n$, or how to reduce the difference between $k_1$ and $k_2$. This is a subject for future work.
\newpage
\bibliography{biblio}
\newpage
\renewcommand{\thesection}{\Roman{section}} 
\renewcommand{\thesubsection}{\thesection.\Roman{subsection}}
\setcounter{section}{0}
\section*{Appendix}
This appendix contains the following:
\begin{enumerate}
\item  {\bf Partial spending from partially validated funds} (Section~\ref{sec:off-chain}): This section includes a description of how partially validated funds can be spent without full validation. It turns out that this is possible if the buyer is correct, but a corrupt buyer can prevent some such payments, but it will be detected. 
\item  {\bf Impossibility Proofs} (Section~\ref{sec:impossible}): shows that no solution to the partial spending problem that tolerates an adaptive adversary can be deterministic or allow the spending of the whole balance.

\item  {\bf Propagating Information} (Section~\ref{sec:propagate}): This section includes the protocols for propagating information and its correctness proof. 

\item  {\bf Partial Spending and Settlement} (Section~\ref{ap:settle}): This section provides code for the buyer settlement protocol and the proofs that our solution for the payment and settlement satisfy the problem requirements.
\item  {\bf $(k_1-k_2)$-Quorum Systems Properties} (Section~\ref{sec:k1k2-proofs}): This section provides the proofs for the lemmas relating to $(k_1-k_2)$-quorum systems that are listed in the main text.
\item  {\bf Quorum Selection Protocol Proofs} (Section~\ref{sec:quorum-select-proofs}): This section contains proofs of the lemmas relating to the quorum selection protocol.
\end{enumerate}

\section{Partial spending from partially validated funds}\label{sec:off-chain}
If a buyer pays a seller with a partial spending transaction, how can the seller turn around and spend some of the received amount without first settling the fund? It turns out, maybe not surprisingly, that in our asynchronous system model with an adaptive adversary, this does not seem to be achievable without some restrictions. One issue is the inability of the seller to prove to a potential buyer that it has a properly validated level-1 payment because that would require divulging the identities of the validators, which is problematic as we described. Also, it is not clear that a zero-knowledge proof of validity is possible. A seller who is the recipient of level-1 payment has only the validation from the buyer (the signatures of the hashes) that it can share as part of a level-2 payment. This would at least prevent a seller who has not received a level-1 payment from attempting to make a level-2 payment, but this is a weak guarantee because both the buyer and the seller can be corrupt.

We outline the main idea for supporting level-$i$ payments, but do not present any protocols. The idea for supporting level-$i$ payments, $i > 1$, is the following. We start with a buyer who has an fully certified fund. The buyer can make up to $k'_2$ level-1 partial spending transactions. If we want to spend from the resulting $k'_2$ funds (level-2 payments), we can do $k'_2$ partial spendings from each one of them resulting in ${k'_2}^2$ level-2 partial spending transactions. So, in general, our approach will allow a total of ${k'_2}^i$ level-$i$ spending transactions, $i \geq 1$. 

Of course all of this should be done so that double spending is not possible.  For level-1 payments, the total spending can be as high as $k'_2 * F.\fbl/k'_2 = F.\fbl$, and potentially there might be no funds left for level-2 payments! One way to deal with this is to reduce the partial spending fraction from $1/k'_2$ to $1/2k'_2$.   This way, the total of level-1 payments is $k'_2 * F.\fbl/2k'_2 = F.\fbl/2$. The total for level-2 payments is ${k'_2}^2 * F.\fbl/4{k'_2}^2 = F.\fbl/4$. So the total spending at all levels is $F.\fbl/2 + F.\fbl/4 + F.\fbl/8 + \ldots \leq F.\fbl$. So, essentially, multi-level spending from partially certified funds can achieved by increasing the amount that cannot be spent from a given balance. Also, it would require that the fund {\em level} be part of the fund description, which could be maintained as a chain of spending transactions starting from level-1 up to the level of the fund. 

One limitation of this approach is that it does not guarantee for every recipient of a level-1 payment the ability to make $k'_2$ level-2 payments. Instead, the guarantee is on the total number of level-2 payments. With this solution, a corrupt buyer, can issue many level-1 payments that are not properly validated to sellers under the control of the adversary. These sellers can then make level-2 payments that {\em consume} the total quota of level-2 payments. So, if the buyer is corrupt, a seller can be denied the opportunity to make level-2 payments, but the buyer's corruption will be detected. The situation can be worse as we go up the levels. Nonetheless, the fact that level-2 and higher payments can be made at all is surprising.

\section{Impossibility Proofs}\label{sec:impossible}
We show that a deterministic solution cannot satisfy all the problem's requirements (Section~\ref{sec:problem}) and that no solution that uses less than full quorums can spend the whole amount and avoid double spending.

\begin{lemma}
If all spending transactions are partially certified (certified by quorums of size less than $f+1$), then a deterministic solution to the spending problem defined in Section~\ref{sec:problem} cannot guarantee progress with \whp for correct buyers and sellers.
\end{lemma}
\begin{proof}
In a deterministic solution, the identities of the validators for a given transaction is solely determined by the buyer and seller identifiers $\pkb$ and $\pks$, the history of transactions at the buyer and seller and the initial states of the buyers and sellers. Consider an execution in which all payments are made by one buyer and assume that the buyer pays the seller with a partial spending transaction. The identity of the validators that will be used by the seller can be calculated by the buyer because it has all the information needed to do that. If the buyer is corrupt, the adversary can learn the identities of the validators for the transaction and corrupts all of them because they number less than $f+1$. The seller can then be prevented from successfully settling the fund resulting from the transaction.
\end{proof}

\begin{lemma}
If all quorums used in validation have size less than $f+1$, then it is not possible to prevent double spending and allow the spending of the whole balance of a fund.
\end{lemma}
\begin{proof}
We consider an execution in which a buyer spends the whole amount in a fund $F$ by executing $k$ partial spending transactions, $T_1,T_2,\ldots,T_k$, for some $k$, to honest sellers who validate and accept the payments. No validators are corrupted during the executions of these $k$ partial spending transactions. After these partial spending transactions are validated, the adversary corrupts the buyer who issues a partial spending transaction $T$ to a corrupt seller. To validate the transaction, less than $f+1$ validators are contacted to validate the transaction. In a general solution, some of the validators for a transaction are contacted directly by the seller and buyer and some of them could be contacted indirectly by other validators. Since the adversary controls the buyer and seller, it knows the identities of the validators that are directly contacted and can corrupt them because there are at most $f$ validators involved in validating the transaction. In turn, and for the same reason, this allows the adversary to learn the identities and corrupt any validators that are used to validate the transaction whether they are directly or indirectly contacted by the seller and buyer. Since all the validators for transaction $T$ can be corrupted by the adversary, the transaction can be successfully validated and successfully settled.   
Since the transactions $T_1,T_2,\ldots,T_i$ are to honest sellers, according to the problem requirements, the sellers for these transactions should be able to successfully settle the funds resulting from them, resulting in a total spending that exceeds the original balance $F$, a double spending.
\end{proof}

\section{Propagating Information: Protocols and Properties}\label{sec:propagate}

As we discussed in the protocol overview, the adversary can 
corrupt a validator if it learns that the validator validated a 
particular transaction. The identities of validators that are involved 
in validation are kept secret by the validators themselves (if honest) or the seller until settlement time, at which time they need to be divulged.  The \textproc{Propagate()} protocol allows a  party (seller or validator) to send secret information to all validators so that the adversary would either
have to corrupt the party to learn the information or, if the adversary learns the information without corrupting the party, then all but $f$ honest are also guaranteed to learn the information. The \textproc{Propagate()} protocol is implemented using secret sharing and reconstruction. The code is given in Algorithms~\ref{propagate-c} and~\ref{propagate-v}.

\begin{algorithm}[t]
    \caption{Propagating Information: Client $c$'s code}
    \label{propagate-c}
    \begin{algorithmic}[1] 
   \Procedure{PropagateClient}{$\mathit{message},N_{prop}$}
     \State $[s_i] = $ \Call{SecretShare}{$\mathit{message}$, $n$, $f$} 
     \For{$i:= 1 \to n$}             
      \State \send $(\mbox{SHARE}, N_{prop},  s_i,\sign_c(s_i||v_i||N_{prop})) \textbf{ to }  v_i$
        \EndFor

\vspace{1ex}
		\State $\mathit{acks}[N_{prop}] = \emptyset$
        \Repeat 
        \If{$\rcvd (\mbox{SHARE\_ACK}, N_{prop}) \textbf{ from } v$}
                \State $\mathit{acks}[N_{prop}] = \mathit{acks}[N_{prop}] \cup  \{v\}$
            \EndIf
        \Until{$|\mathit{acks}[N_{prop}]| \geq n-f$}
        \State \send $(\mbox{RECONSTRUCT}, N_{prop}) \textbf{ to } \validators$ 
        
        \vspace{1ex}
        \State $\mathit{reconstructed}[N_{prop}] = \emptyset$
        \Repeat 
        \If{$\rcvd (\mbox{RECONSTRUCTED}, \mathit{message}, N_{prop}) \textbf{ from } v$}
                \State $\mathit{reconstructed}[N_{prop}] = \mathit{reconstructed}[N_{prop}] \cup  \{v\}$
            \EndIf
        \Until{$|\mathit{reconstructed}[N_{prop}]| \geq n-f$}
        
       \EndProcedure
    \end{algorithmic}
\end{algorithm}

\begin{algorithm}[t]
    \caption{Propagating Information: validator $v$'s code. The client $c$ in the code can be either a seller invoking the client side of information propagation to settle a fund resulting from partial payment or a validator invoking propagation information to handle a buyer's settlement transaction}\label{propagate-v}
    \begin{algorithmic}[1] 
        \Procedure{PropagateServer}{$N_{prop}$}
        \State {\bf repeat}
        \State\hspace{3ex} \upon \rcpt $(\mbox{SHARE},N_{prop}, s,\sigma = \sign_\pks(s||v||N_{prop})) \textbf{ from } c$: 
			\State\hspace{6ex} \send $(\mbox{SHARE\_ACK}, N_{prop}) \textbf{ to } c$
			\State\hspace{6ex} $\mathit{share}[c,N_{prop}] = s,\sigma$
     
        \vspace{1ex}
        \State\hspace{3ex} \upon \rcpt $(\mbox{RECONSTRUCT}, N_{prop}) \textbf{ from } c$:
     		 \State\hspace{6ex} \send $(\mbox{FORWARD}, c, \mathit{share}[c,N_{prop}], N_{prop}) \textbf{ to } \validators$
		     \State\hspace{6ex} $\mathit{shares}[c,N_{prop}] = \emptyset$, $\mathit{Forwarded}[c,N_{prop}] = \emptyset$

	             \vspace{1ex}
	   			
			\vspace{1ex}
        \State\hspace{3ex} \upon \rcpt $(\mbox{FORWARD},c,s, \sigma,  N_{prop})$ \textbf{ from } $v$:
       
        	\State\hspace{6ex} \bif $\sigma = \sign_c(s||v||N_{prop}) \wedge v \notin \mathit{Forwarded}[c,N_{prop}]$ {\bf then}

                \State\hspace{9ex} $\mathit{shares}[c,N_{prop}] = \mathit{shares}[c,N_{prop}] \cup  \{ (s,\sigma)\}$                				
                \State\hspace{9ex} $\mathit{Forwarded}[c,N_{prop}] = \mathit{Forwarded}[c,N_{prop}] \cup  \{ v \}$
        \State \hspace{9ex} \bif {$|\mathit{shares}[c,N_{prop}]| \geq f+1$} {\bf then}
			 \State\hspace{12ex} $\mathit{message}[c,N_{prop}] =$ 
			 				\Call{ReconstructSecret}{$\mathit{shares}[c,N_{prop}]$}
         	\State\hspace{12ex} $\send (\mbox{RECONSTRUCTED}, \mathit{message}[c,N_{prop}], 
								N_{prop}) \textbf{ to }  c \mbox{ and } \validators$ 

       \State {\bf until} $(n-f)$ RECONSTRUCT messages received
       \EndProcedure
    \end{algorithmic}
\end{algorithm}

Propagating information is relatively simple in our setting and is easier than the asynchronous verifiable secret sharing problem~\cite{cachin2002asynchronous} (a solution to which can also be used, but would be an overkill). The client protocol takes as input two arguments: the {\em message} being propagated and a unique nonce $N_{prop}$ to distinguish between different calls to the propagate protocol. The client $c$ who propagates the information creates shares of the information
that needs to be propagated so that any $f+1$ out of $n$ shares can be used to reconstruct the {\em message}. Depending on the settlement protocol, $c$ is either a seller or a validator.  Each share $s_i$ is addressed to a particular validator $v_i$. In addition to the share $s_i$, $v_i$ receives from the client a signature $\sigma_i = \sign_c(s_i||v_i||N_{prop})$ that $v_i$ can later use to prove that $s_i$ is a properly received share. After the shares are distributed and $n-f > f+1$ validators acknowledge receiving the shares,
the client sends a RECONSTRUCT message so that the validators can reconstruct the {\em message} from the shares. Validators broadcast their shares and collect shares that they authenticate until $f+1$ authenticated shares are collected. At that point, the {\em message} is reconstructed using the collected shares. It is important to note here that all the shares used in the reconstruction are validated  to be form the client (line 9 of validator code), so if the client is honest, the $f+1$ shares will all be valid even if they are received from corrupt validators. If the client is not honest, we don't care about the value that is reconstructed.
A validator stops participating in the propagation protocol for a particular $N_{prop}$ when $n-f$ different validators announce that they have reconstructed the messages. Of those validators, $n-2f$ must be honest.

The algorithms guarantees that if the client calling \textproc{Propagate($\mathit{message}$)} is honest, all but $2f$ honest validators will receive $\mathit{message}$. If the client propagating the shares is not honest, there are no guarantees as to what is reconstructed and the reconstruction might even fail, but that does not affect the algorithms that use the \textproc{Propagate()} function.

\begin{lemma}
If an honest client calls \textproc{Propagate($m$)}, all but $2f$ honest validators will be guaranteed to receive $m$. 
\end{lemma}
\begin{proof}(sketch)
If the client is honest, when a validator receives a RECONSTRUCT message, $n-f$ validators must have already received the SHARE message. Of these validators $n-2f \geq f+1$ validators are honest and will each send FORWARD message containing its share to every other honest validator who will be able to reconstruct the message.
Validators stop their execution when they receive $n-f$ RECONSTRUCT messages from other validators. Of these $n-2f$ are from honest validators.
\end{proof}

\begin{lemma}
Let $N$ be a random value that is generated by a client $c$ such that no value that is directly or indirectly dependent on $N$ is sent to any validator with the exception of  $\textproc{H}(N)$. If the adversary learns $N$ \whp, then \whp the adversary must have corrupted the client $c$.
\end{lemma}
\begin{proof}(sketch)
The proof is by induction on the number of messages sent by $c$. If $c$ sends no messages, it should be clear that the adversary cannot learn $N$ without corrupting $c$. Assume that the lemma holds for the first $i$ messages sent by $c$ and consider the $i+1$ message sent by $c$. The $i+1$ message either contains no value dependent on $N$ or contains $\textproc{H}(N)$ and other values that are not dependent on $N$. In the first case, the adversary clearly doesn't learn anything about $N$. In the second case, given the adversary's bounded computational power, it cannot learn the value of $N$ except with negligible probability. It follows that the chain must be a 0-length chain and the adversary learns $N$ by corrupting $c$.
\end{proof}

\begin{lemma}
Let $N$ be a random value that is generated by a client $c$ such that no value that is directly or indirectly dependent on $N$ is sent to any validator with the exception of  $\textproc{H}(N)$ or by executing \textproc{Propagate}($m$)for a message $m$ that depends on $N$. If the adversary learns $N$ \whp, then \whp the adversary must have corrupted the client $c$ or all but $2f$ honest validators are guaranteed to learn $m$.
\end{lemma}
\begin{proof}(sketch)
Since each share generated by the secret sharing of the \textproc{Propagate}($m$) is independent of $N$, by an argument similar to that in the previous lemma, the adversary cannot learn $N$ \whp if it does not learn $f+1$ shares. If the adversary learns $N$, then it must have corrupted a validator  that received $f+1$ valid shares. One of these shares must be from an honest validator that received a RECONSTRUCT message. It follows that $n-f$ validators must have received a reconstruct message and all of these validators send FORWARD messages to all other validators who will be able to reconstruct $m$. Validators participate in the protocol until $n-f$ validators announce that they reconstructed the message at which point $n-2f$ honest validators must have reconstructed the message.
\end{proof}

It might seem a little strange that even though every honest validator sends messages to every other honest validator, we only guarantee that $n-2f$ validators will receive the message if the adversary doesn't corrupt the client. The reason is that even though honest validators might be guaranteed to receive messages sent by other honest validators eventually, we need to make a statement about what holds when the execution of \textproc{Propagate}($m$) ends.

\section{Partial Spending and Settlement Proofs}\label{ap:settle}

The protocol for buyer fund settlement is shown in Algorithms~\ref{buyer-settlement}
and~\ref{v-buyer-settlement}. The rest of this section presents the correctness proof of the solution for partial spending and settlement protocols.

\begin{algorithm}[t]
    \caption{Buyer Fund Settlement}
    \label{buyer-settlement}
  \begin{algorithmic}[1] 
    \Procedure{BuyerSettle}{$F$}
        \State \send $(\mbox{SETTLE},F) \textbf{ to }  \validators$
         \Repeat 
            \If{$\rcvd \mathit{resp} \textbf{ from } q_i$} \color{blue}\Comment{Wait for replies from }\color{black}
                \State $\mathit{replies} = \mathit{replies} \cup \mathit{resp} $ \color{blue}\Comment{validators until enough }\color{black}
            \EndIf
        \Until{$\exists F'\,:\,|\{\mathit{resp}\,:\, \mathit{resp}\in \mathit{replies} \wedge \mathit{resp} = F'\}| = n-2f$} \color{blue}\Comment{identical replies are received}\color{black}

        \State \textbf{return } $F'$
    \EndProcedure
  \end{algorithmic}
\end{algorithm}

\begin{algorithm}[t]
    \caption{Code of Validator $v$ for Buyer Fund Settlement}
    \label{v-buyer-settlement}
  \begin{algorithmic}[1] 
    \Procedure{ValidatorBuyerSettle}{$F$, $\pkb$}
    \If{$\rcvd (\mbox{SETTLE},F) \textbf{ from } 
    		\pkb \wedge \pkb \in F.\owners$} 
    	\State $N_{prop} \leftarrow \{0,1\}^n$	
				\color{blue}
				\Comment{different validators choose different $N_{prop}$ values} 
				\color{black}
        \State $value[F,N_{prop}] = \bot$ 
        		\color{blue}
				\Comment{$F$ will have multiple entries, one for each such value} 
				\color{black}

        \vspace{1ex}
        \color{blue}\Comment{If $v$ is witness to payment from $F$, propagate payment info} \color{black}
        \If{$\exists \,(\tid, h_s,\sigma, N) \in \mathit{validated\_transactions}$ $\wedge$
        	$\tid.F = F$ $\wedge$ and $pkb \in \tid.\owners$} 
          \State  \Call{Propagate}{$\langle \langle F,\mbox{SETTLE}\rangle \rangle$, $v$ , 
          			$(\tid,h_s,\sigma, N), N_{prop}$}

        \Else \color{blue}\Comment{otherwise propagate that $v$ is not witness to any payment from $F$} \color{black}
          \State \Call{Propagate}{$\langle \langle F,\mbox{SETTLE}\rangle \rangle$, $v$ , 
          	$\sign_v(F||none),N_{prop}$} 
	        \EndIf
    \EndIf

    \vspace{1ex}
    	\color{blue}\Comment{Update } known $\mathit{transactions}[F]$ based on propagated information\color{black}
    \If{$\mathit{message}[F,N_{prop}] = \langle \langle F,\mbox{SETTLE}\rangle \rangle$, $v$ , 
          			$(\tid,h_s,\sigma, N), N_{prop}\rangle$}
\State $\mathit{transactions}[F] = \mathit{transactions}[F] \cup \{(\tid,h_s)\}$
     \EndIf
     
	\vspace{1ex}
	\color{blue}\Comment{Forward any witness or non-witness information received} \color{black}
    \If{$\mathit{message}[F,N_{prop}] = (\langle F,\mbox{SETTLE}\rangle, v', *)$}
       \State $\send (F, \mbox{SETTLE}, value[F,N_{prop}] )$ to all validators
     \EndIf

	\vspace{1ex}
    \If{$\rcvd (\langle F,\mbox{SETTLE}\rangle, v', *)$ 	
		$\wedge$ $v' \not\in \mathit{settle}\_\mathit{validators}[F]$}
         \State $\mathit{settle}\_\mathit{validators}[F] = 
         		\mathit{settle}\_\mathit{validators}[F] \cup v'$
    \EndIf
     
    \vspace{1ex}   
    \If{$\rcvd (F, \mbox{SETTLE}, v' , (\tid,\sigma,N))$ 
    		$\wedge$ $v' \not\in \mathit{settle}\_\mathit{validators}[F]$ $\wedge$
		
		\hspace{1ex} 
			$\Call{ValidPayment}{F, (\tid,h_s,\sigma,N})$}                   
			\State $\mathit{payments}[F] = \mathit{payments}[F] \cup 
											(\tid, h_s,\sigma, N)$
    \EndIf

       	\vspace{1ex}
    \If{$|\mathit{settle}\_\mathit{validators}[F]| = n-f$}
                \ForAll{$(\tid, h_s,\sigma, N) \in \mathit{payments}[F]\,:\, 	
                		\Call{ValidPayment}{\tid,h_s,\sigma,N}$}
                \State $\mathit{transactions}[F] = \mathit{transactions}[F] \cup \{(\tid,h_s)\}$
               
                \State $\mathit{settle} = \mathit{settle} \cup \{F\}$ 
                \If{$|\mathit{transactions}[F]| > k_1$}        		\color{blue}
				\Comment{if many transactions were issued from $F$} 
				\color{black}
                	\State {\bf abort} \color{blue}
				\Comment{abort} 
				\color{black}
				\EndIf
        		\color{blue}
		
		\vspace{1ex}
				\Comment{Add $F$ to funds with SETTLE transactions} 
				\color{black}
				 \EndFor
          		\State $\mathit{new}\_\mathit{balance} = F.\fbl - |\mathit{transactions}[F]|\times 1/k'_2$
         		\State $F'.\tid = H(F.\tid)\,;\, F'.\fbl = \mathit{new}\_\mathit{balance}\,;\, F'.\owners = F.\owners$
        \State $\send (F',\sign_v(F')) \textbf{ to } F'.\owners$
     \EndIf
    \EndProcedure
  \end{algorithmic}
\end{algorithm}

We show that the protocol satisfies the problem requirements.

\begin{lemma}[Transactions to honest sellers are accounted for]
Let $T_{s} = (\tx, N_s)$ be a partial spending transactions for which the seller is honest. If the buyer executes $(\mbox{SETTLE},F)$, then $n-2f$ honest validators will have $(\tx, h_s = \textproc{H}(N_s)) \in \mathit{transactions}[F]$. 
\end{lemma}
\begin{proof}(sketch)
If the seller already executed a settlement transaction for the fund resulting from $T_s$, then it should have received validations from $n-f$ validators each of which adds $(\tx, h_s = \textproc{H}(N_s))$ to $\mathit{transactions}[F]$. So, we assume that the seller did not execute a settlement transaction for the fund resulting from $T_s$ and did not divulge $N_s$. It follows that the identities of the validators of $T_s$ are not known to the adversary which means, by the model assumptions, that some of the honest validators of $T_s$ will succeed in propagating their information to $n-2f$ honest  validators that add 
$(\tx, h_s)$ to $\mathit{transactions}[F]$. 
\end{proof}

\begin{lemma} All partially certified funds with honest owners can be settled successfully. 
\end{lemma}
\begin{proof}(sketch)
Consider a partially certified fund $F_s$ resulting from a partial spending transaction $\tid$ from a fund $F$. If the owner (seller) of $F_s$ is honest, then when the payment was made, the seller selected a quorum according to the quorum selection protocol and then got the payment validated by a sufficient number of honest validators from the selected quorum. The identities of these validators are not known to the adversary before the owner settles the fund. Also, each of these honest validator received the validation request from the seller before receiving a settlement request for $F$ because one of the conditions for validating a payment request is for the validator not to have $F$ is the set of funds for which there is a SETTLE transaction ($F.\tid \not\in \mathit{settle}$). Every honest validator that receives the settlement request for $F_s$ will either have $F \not\in \mathit{settle}$ or  $F \in \mathit{settle}$. If $F \in \mathit{settle}$, by the previous lemma, the validator must have added  $F_s$ to $\mathit{transactions}[F]$.
The validation of the settlement of $F_s$ will go through in both cases because the other conditions for validating a seller's settlement transaction will hold because the honest seller provides a correctly selected quorum and a large enough set of witnesses that validates the partial spending transactions that created $F_s$.
\end{proof}

\begin{lemma}
Settlement for a partially certified funds equals payment amount: If $F''$ is a fully certified fund resulting from executing $(F',\mbox{SETTLE})$ for partially validated fund $F'$, then $F''.\fbl = F'.\fbl$.  
\end{lemma}
\begin{proof}(sketch)
This follows immediately from the code. When a fund $F'$ of a partial spending transaction is settled, the balance of the resulting  fund $F''$ is equal to $F.\fbl/k'_2$ which is also the spending amount (which we do not explicitly represent in the protocols). 
\end{proof}

\begin{lemma}
Settlement amounts for payments from $F$ are subtracted from settlement for $F$:  If executing transaction $(F,\mbox{SETTLE})$ results in a fully certified fund $F_R$:

   \[  F_R.\fbl \leq  F.\fbl - \sum_{F' \in \settled_F} F'.\fbl
    \] 
\end{lemma}
\begin{proof}(sketch)
Consider $F' \in  \settled_F$ that results from settling a fund $F_s$ identified by  
transaction $(\tid,N_s)$ when $F_s$ is settled, resulting in $F'$, either $F$ is not yet settled and $(\tid,\textproc{H}(N_s))$ gets added to $\mathit{transactions}[F]$ by the validators of the settlement transaction   or  $(\tid,\textproc{H}(N_s))$ is already in $\mathit{transactions}[F]$. In either case, when 
\end{proof}

\begin{lemma}[No more than $k_2+f/v_s$ partial spending transactions]
With high probability, no more than $k_2+f/v_s$ partial spending transaction can be validated, where $v_s$ is the validation slack.
\end{lemma}
\begin{proof}
With high probability, $k_2$ is an upper bound on the number of partial spendings that can be validated without any failures. If the adversary can only randomly chose the validators to corrupt, which is the case when the seller is honest, then $k_2$ will also be the number of partial spendings that are possible. So, we assume that $k_2$ partial spending transactions are executed with no validator corruption. Any additional transactions would have quorums of validators that intersect the previously selected quorums in $\beta m$ validators and the adversary would need to corrupt $(\beta-\alpha)m$ for each additional partial spending transactions so that the number of validators that validate the transaction is raised to $\alpha m$. A total of $f/((\beta-\alpha)m$ additional spending transactions. The total number of partial spending transactions possible is therefore $k_2 +f/((\beta-\alpha)m$ =  $k_2+f/v_s$. 
\end{proof}
    
\begin{lemma}
If the owner of a fund $F$ is honest, settlement amount is no less than the the initial balance of $F$ minus payments made from $F$: 
    If executing transaction $(F,\mbox{SETTLE})$ results in a fully certified fund $F_R$ and $F.\owners$ is honest:
     \[  F_R.\fbl \geq F.\fbl - \sum_{F' \in \funds_F} F'.\fbl
    \]
\end{lemma}
\begin{proof}(sketch)
This follows directly from how the settlement amount is calculated in which only transactions originating from the owner of $F$ are deducted from the resulting balance.
\end{proof}

\begin{lemma}[Non-interference] If a total of $k \leq k_1$ payment transactions are initiated by $\pkb \in F.\owners$ from  fully certified fund $F$ and no additional payment or settlement transactions are initiated by $F.\owners$ and $\pkb$ is honest, then,  every one of the $k$ transactions whose seller (payee) is honest will be validated.\end{lemma}
\begin{proof}(sketch)
We consider one of these transactions with an honest seller. The seller will select a quorum of validators according to the protocol and gets validations from the buyer. Then the seller will contact the validators to get the transaction validated. By the properties of $(k_1,k_2)$-quorum systems, the seller will get replies from $m - (1+\mu) p_f m$ validators, $(1 - \alpha) \times m$ of which have not previously validated another transaction from $F$ and the transaction will be validated.
\end{proof}

\begin{lemma}
Successful settlement for fully certified funds: If the owner of fully certified fund $F$ is honest and executes an $(F,\mbox{SETTLE})$ transaction, the settlement transaction for $F$ will terminate. 
\end{lemma}
\proof{lemma}
Since the owner is honest, at most $k_1$ partial spending transactions from $F$ can be executed. When the $(F,\mbox{SETTLE})$ transaction is executed, the validators will only have no more than $k_1$ transactions in $\mathit{transaction}[F]$ because every transaction in $\mathit{transaction}[F]$ must be checked for validity with 
\textproc{ValidPayment}($\tid,h_s,\sigma,N$) which checks that sigma is a valid signature by the buyer for $\tid||h_s||N$. Since the only potentially blocking condition in the validator's buyer settlement code is when there are more than $k_1$ transactions in $\mathit{transaction}[F]$, every honest validator that handles the settlement will validate $(F,\mbox{SETTLE})$ and eventually the buyer will get $n-2f$ identical replies from validators. 
\proof{lemma}

\section{\texorpdfstring{$(k_{1},k_{2})$}--Quorums Properties}\label{sec:k1k2-proofs}

\begin{L1}
An $(m,n)$ uniform balanced quorum system is a $(k_1,k_2)$-quorum system with $\alpha = 1/3$ and $\beta = 2/3$ has $\epsilon = \delta = \negl[m]$ and {\em validation slack} = $m/3$, if $p_f+\alpha_1 < 1/3$, where $\alpha_1 = k_1m/n$ and $p_f = f/n$.
\end{L1}
\begin{proof}
We need to show that:

\begin{enumerate}
    \item Lower bound: $\Pr_{Q , Q_j \leftarrow  \mathcal{Q}\,;\, j \in J_1; \,|J_1| \leq k_1}[|Q \cap ({\mathcal{F}_{pr}}\cup \bigcup_{j \in J_1} Q_j)| > m/3 ] \geq 1 - \negl[m]$
    \item Upper bound: $\Pr_{Q , Q_j \leftarrow  \mathcal{Q}\,;\, j \in J_2; \,|J_2| \geq k_2}[|(Q \cap (\bigcup_{j \in J_2} Q_j))-{\mathcal{F}_{pr}}| \leq 2m/3 ] \geq 1 - \negl[m]$
\end{enumerate}

For the lower bound, consider the intersection of a quorum $Q$ with the union of up to $k_1$ previously selected quorums. Since $k_1m/n = \alpha_1$, $\alpha_1 n$ is an upper bound on the size of the union of the $k_1$ previously selected quorums because each of them is of size $m$. The expected size of the intersection of $Q$ with these $k_1$ previously selected quorums is at most $\alpha_1|Q| = \alpha_1 m$ because the probability that a given randomly chosen server in $Q$ is in the union is at most $\alpha_1 n / n = \alpha_1$. Similarly, the probability that a server in $Q$ is previously corrupted is at most $p_f$ and the expected number of servers in $Q$ that are previously corrupted is at most $p_f|Q| = p_fm$. So, the expected size of the intersection of $Q$ with the union of the previously corrupted servers together with the union of $k_1$ previously selected quorums is at most $(\alpha_1+p_f)m$. Since $\alpha_1+p_f < 1/3$, $(\alpha_1+p_f)m < m/3$. Let $r = (1/3(\alpha_1+p_f))-1$. This value of $r$ is positive and independent of $m$. For this value of $r$, we have $(1+r)\times (\alpha_1+p_f)m = m/3$. So, the probability that the size of the intersection exceeds by a factor $1+r$ the expected size of the intersection is the same as the probability that the size of the intersection exceeds $m/3$

By Chernoff's upper tail bounds~\cite{probability-book}, the probability that the size of the intersection exceeds by a factor of $1+r$, $r > 0$, the upper bound $(\alpha_1+p_f)m$ on the expected size is:

\[
\Pr_{Q \leftarrow  \mathcal{Q}}[|Q \cap ({\mathcal{F}_{pr}}\cup \bigcup_{j \in J_1} Q_j)| \geq (1+r)\times (\alpha_1+p_f)m = m/3] \leq e^{-r^2 \times (\alpha_1+p_f)m/3 } 
\]
which is a negligible function of $m$. Now, it remains to show that the upper bound requirement holds: \[\Pr_{Q , Q_j \leftarrow  \mathcal{Q}\,;\, j \in J; \,|J| \geq k_2}[|(Q \cap (\bigcup_{j \in J} Q_j))-{\mathcal{F}_{pr}}| < 2m/3 ] \geq 1 - \negl[m]
\]

The expected number of servers in a quorum $Q$ that are in the union of $k_2$ previously selected quorums is $(k_2m/n)|Q| = (1-\alpha_1)m$. The expected number of servers in $Q$ that are previously corrupted is less than $p_f m$. So, the expected number of non-corrupted servers in $Q$ that are in one of the last $k_2$ chosen quorums is at least $(1-\alpha_1)m - p_fm = (1 -(\alpha_1+p_f))m > 2m/3$ because $\alpha_1+p_f < 1/3$ as we have noted above. 
Again, if we define $r = 1 - ((1 -(\alpha_1+p_f))/(2/3)$, we have $0 < r < 1$, and $(1- r) \times (1 -(\alpha_1+p_f))m = 2m/3$.
Using Chernoff's lower tail bound, we have 
\[\Pr_{Q , Q_j \leftarrow  \mathcal{Q}\,;\, j \in J_2; \,|J_2| \geq k_2}[|(Q \cap (\bigcup_{j \in J} Q_j))-{\mathcal{F}_{pr}}| \leq (1- r) \times (1 -(\alpha_1+p_f))m = 2m/3] \leq e^{-r^2 (1 -(\alpha_1+p_f))m/2}\]
which is a negligible function of $m$. It is important to note that $r$ does not depend on $m$ but on the value of $\alpha_1+p_f$   which is a constant for a given $p_f$ and $k_1$ and $k_2$. So, the function above decrease exponentially with $m$ for fixed $\alpha_1$ and $p_f$ satisfying $\alpha_1+p_f < 1/3$. Also, for a fixed $k_1$ and $\alpha_1$, $m$ grows linearly with $n$: $m = \alpha_1 n/k_1$. In other words, the probabilities are negligible functions of $n$ for fixed $k_1$, $\alpha_1$ and $p_f$.
\end{proof}

Even though the proof shows that the probability of not satisfying the $(k_1,k_2)$-quorum requirements is negligible, values of $\alpha_1$ and $p_1$ for which the sum $\alpha_1+p_f$ is very close to 1/3 result in probability bounds that are not useful in practice. For the parameter choices we made, the {\em validation slack} of the system is $m/3$. Different values for $\alpha_1$ and $p_f$ could result in lower validation slack but better overall performance. Determining the optimal combination of parameters is subject of future work.
\begin{L2}
An $(m,n)$ uniform balanced quorum system is a $(k_1,k_2)$-quorum system is a $(k_1,k_2, \epsilon, \delta, \alpha = 1/3, \beta = 2/3, \mu )$-asynchronous quorum system with $\epsilon = \delta = \negl[m]$ and {\em validation slack} = $m/3$, if $n > 8f$ and $k_1m/n < 1/24$.
\end{L2}
\begin{proof}
The proof is almost identical to that of Lemma~\ref{lem:uniform-prob}. Recall that for asynchronous $(k_1,k_2)$-quorum systems, the properties intersection and non-intersection properties should hold even if we exclude from a quorum $Q$ a set $Q_s$ of size at most $(1+\mu)p_fm$.
For the non-intersection property, not including some replies can only make the intersection smaller, so the proof that with negligible probability, the intersection size is less than $m/3$ carries over to this setting. For the upper bound, the only difference is that we are excluding an additional $(1+\mu)p_f n$ servers in addition to the already excluded $p_f n$ servers. The expected size of non-corrupt servers in $Q$ that are also in one of the previous $k_2$ chosen quorums is at least $(1-\alpha_1 - p_f)m - (1+\mu)p_f m = (1-\alpha_1 - (2+ \mu) p_f)m$. We show that the probability that the size is short by a factor $(1-r)$ of the expected size is negligible, where .  
 $r = 1 - ((1 -(\alpha_1+p_f+(1+\mu)p_f))/(2/3)$. Similarly to what we did in Lemma~\ref{lem:uniform-prob} for the synchronous case, we have $0 < r < 1$ and 
using Chernoff's lower tail bound, we have for any $Q_s$ such that  $|Q_s| \leq m p_f (1+\mu)$:
\begin{align}
 &\Pr_{Q , Q_j \leftarrow  \mathcal{Q}\,;\, j \in J_2; \,|J_2| \geq k_2}[|((Q-Q_s) \cap ((\bigcup_{j \in J}  Q_j)-{\mathcal{F}_{pr}})| < 2m/3] \\
& =  \Pr_{Q , Q_j \leftarrow  \mathcal{Q}\,;\, j \in J_2; \,|J_2| \geq k_2}[|(Q \cap ((\bigcup_{j \in J} Q_j)-{\mathcal{F}_{pr}-Q_s)}|] < \leq (1- r) \times (1 -(\alpha_1+p_f)m] \\
&  \leq e^{-r^2/2 \times (1 -(\alpha_1+p_f)m}
\end{align}
\end{proof}

\section{Random Quorum Selection}\label{sec:quorum-select-proofs}

The algorithm for random quorum selection is shown in Algorithm~\ref{alg:quorum-select}. The following lemmas establish the relevant properties of the algorithm.

\begin{algorithm}[!t]
    \caption{Random selection of a quorum of size $m$: Seller's's Code ($\mathit{pks}$)}
    \label{random-quorum-selection}
    \begin{algorithmic}[1] 
        \Function{SelectQuorum}{$\langle F, \pkb , \pks \rangle , N_s$}

    \State $T_s = \langle F, \pkb , \pks \rangle ||N_s$ \color{blue}\Comment{Seller's transaction identifier}\color{black}\label{Tunique}
       \State $h = H(T_s)$ 
     \State  $Q = \{ \}$ ; $j = 1$ ; 
     \While{$|Q|< m$}
           \If{$\mathit{Server}(H(h||j)) \not\in Q$} \color{blue}\Comment{If server not previously selected,}\color{black}\label{line:star}
        \State $Q = Q \cup \{\mathit{Server}(H(h||j))\}$ \color{blue}\Comment{add server to quorum}\color{black}
              \EndIf
        \State $j = j+1$
     \EndWhile
     \State \Return $Q$ 
	           \EndFunction
    \end{algorithmic}\label{alg:quorum-select}
\end{algorithm}

\begin{L3}
If the seller is correct, the quorum of validators is chosen uniformly at random.
\end{L3}
\begin{proof}
If the seller is correct, the validators are chosen using the hash function on different input values, seeded with a high entropy $N_s$, that \whp have not been previously selected, until $m$ different validators are chosen. Given that the hash function is modeled as a random function, each of these validators will be chosen randomly.
\end{proof}

\begin{L4}
If the seller is correct, the identities of the correct validators in the chosen quorum are not known to the adversary or to the buyer.
\end{L4}
\begin{proof}
Under the assumption that the adversary cannot intercept the messages of the seller relating to one transaction (Section~\ref{sec:model}), the adversary can only learn about the transaction from corrupt servers that are selected as part of the quorum or by guessing the seed. Given that the seed has high entropy, it cannot be guessed other than with negligible probability. The rest of the proof follows directly from the fact that the validators are randomly chosen. All validators that are not under the control of the adversary are equally likely to be chosen in a particular quorum.
\end{proof}

\begin{lemma}\label{lem:corrupt-validator-ratio}
The probability that a randomly selected quorum of size $m$ contains $(1 + \mu)p_fm$ previously corrupt validators, $0 < \mu < 1$ is at most $e^{-\mu^2 \times p_f m/(2+\mu)}$.
\end{lemma}
\begin{proof}
Let $p'_f$ be the fraction of validators that have been previously corrupted. $p'_f < p_f$.
Let $\mathcal{Q}_m$ be the set of subsets of validators of size $m$. The probability that a randomly chosen validator is corrupt is at most $p'_f$. The expected number of validators that are corrupt in a quorum of size $m$ is at most $p'_fm$. By Chernoff's bounds, we have 
\[
\Pr_{Q \leftarrow  \mathcal{Q}_m}[|Q \cap {\mathcal F}| \geq (1+\mu) p'_f m] \leq e^{-\mu^2 \times p'_fm/(2 + \mu)} \leq e^{-\mu^2 \times p_fm/(2 + \mu)}.
\]
\end{proof}

The following Lemma follows directly from Lemma~\ref{lem:corrupt-validator-ratio}. 

\begin{L5}\label{cor:bound-on-corrupt}
For $0 < \mu < 1$, if a seller choses $K$ quorums, of size $m$ each, at random, then with probability at most $Ke^{-\mu^2 \times p_f m/(2+\mu)}$, every chosen quorums has no more than $(1+\mu)p_fm$ previously corrupt validators.
\end{L5}

\end{document}